\documentclass[twocolumn,australian,english,aps, prl]{revtex4-1}
\usepackage[T1]{fontenc}
\usepackage[latin9]{inputenc}
\usepackage{color}
\usepackage{bm}
\usepackage{amsmath}
\usepackage{amssymb}
\usepackage{graphicx}
\usepackage{babel}
\begin{document}
\title{Hidden causal loops, macroscopic realism and Einstein-Podolsky-Rosen-Bell
nonlocality: forward-backward stochastic simulations }
\author{M. D. Reid and P. D. Drummond}
\affiliation{Centre for Quantum Science and Technology Theory, Swinburne University
of Technology, Melbourne 3122, Australia}
\begin{abstract}
We analyze quantum measurement and entanglement by solving the dynamics
of stochastic amplitudes that propagate both forward and backward
in time. The model allows simulation of Einstein-Podolsky-Rosen and
Bell correlations, and reveals consistency with a weak form of local
realism defined after the unitary interactions determining the measurement
settings. Bell violations emerge due to a breakdown of a subset of
Bell's local-realism conditions. Our results elucidate how hidden
causal loops can explain Bell nonlocality, without requiring retrocausality
at a macroscopic level.
\end{abstract}
\maketitle
The causal structure associated with violations of Bell inequalities
remains a mystery. Bell proved the incompatibility between local causality
and quantum mechanics \citep{bell-1964,bell-2004,clauser-shimony-1978},
motivating the question of whether superluminal disturbances were
possible and hence leading to no-signaling theorems \citep{eberhard-1989}.
However, Bell violations can arise from classical field models, using
retrocausal solutions from absorber theory \citep{pegg-1980}. This
has inspired much research into causality in quantum mechanics \citep{araujo-2015,barrett-2021,wharton-2020,h-price-2008,scully1982,giarmatzi-2019,costa-2016,drummond-2019,hall-2020,wood-2015,hossenfelder-2020,donadi-hossenfelder,time-order}.

While it is possible to account for Bell nonlocality using retrocausality
\citep{h-price-2008,pegg-1980,wharton-2020}, it is not apparent why
retrocausality does not manifest at a macroscopic level. Moreover,
classical causal models involving retrocausality or superluminal signalling
require a ``fine-tuning'' of parameters to explain no-signalling
consistently with nonlocality \citep{wood-2015}, which restricts
the type of model that could be used \citep{finetuning,wharton-2020}.
If Bell nonlocality has a retrocausal origin, it has been suggested
this may involve hidden causal loops \citep{castagnoli-2021}. Recent
work has shown that cyclic causation is not ruled out by no-signaling
restrictions \citep{vilasini-2021}. 

Causal models are connected with the concept of realism. Einstein,
Podolsky and Rosen (EPR) \citep{epr-1935} referred to ``elements
of reality'' as values predetermining a measurement outcome. EPR
also assumed \emph{locality},  \textcolor{black}{that there is no
disturbance to one system due to a space-like separated measurement
on another.} While Bell violations falsify EPR's local realism, it
is not clear whether this is due to failure of locality or realism
(or both) \citep{bell-1964,bell-2004,clauser-shimony-1978}. It
has also been unclear whether, or how, realism holds macroscopically
\citep{leggett-1985,schrodinger-1935}. Macroscopic realism (MR) posits
that for a system in a superposition of macroscopically distinct states,
there is a predetermined outcome $\widetilde{\lambda}$ for a measurement
indicating which of the states ``the system is in'' \citep{leggett-1985}.
MR provides a foundation for macroscopic causality, since the variable
$\widetilde{\lambda}$ gives a value for the outcome of the measurement
at a certain time (independent of future or spacelike-separated events),
thereby avoiding retrocausality at a macroscopic level.

In this Letter we prove a connection between Bell nonlocality and
causal loops involving  ``hidden'' amplitudes for a stochastic
model arising from quantum field theory, whose dynamics has initial
and final boundary values. The solutions are based on phase-space
theorems derived in a companion paper \citep{companion-paper} that
lead to simulations of EPR and Bell correlations. A consequence is
a causal structure arising from a quantum treatment\emph{ }that gives
both backward- and forward-propagating stochastic amplitudes, and
that is consistent with macroscopic realism. Our solutions identify
weak forms of  both locality and realism that are valid in the presence
of Bell violations, with the violations arising from a failure of
an identifiable subset of EPR and Bell's local-realistic conditions.
We show that one can have more general ``beables'' as suggested
by Bell \citep{bell-2004,beables} but with different dynamical properties
to those assumed by EPR and Bell.

To explain how macroscopic realism may be compatible with a microscopic
retrocausal theory, we solve for the dynamics of ``hidden variables''
that are trajectories $\bm{\lambda}\left(t\right)$ for phase-space
amplitudes \citep{drummond-2020,simon-phi}. A quantum state is equivalent
to a positive phase-space probability distribution $Q\left(\bm{\alpha},t\right)=\frac{1}{\pi^{M}}\left\langle \bm{\alpha}\right|\rho\left(t\right)\left|\bm{\alpha}\right\rangle $,
for an M-mode state with density matrix $\rho$ \citep{husimi-1940}.
Real phase-space coordinates $\bm{\lambda}=(\bm{x},\bm{p})$ are related
to complex amplitudes by $\bm{\alpha}=(\bm{x}+i\bm{p})/2$. The operators
that correspond to $\bm{\lambda}$ are complementary, with $\left[\hat{x},\hat{p}\right]=2i$.
The trajectory dynamics of the $\bm{\lambda}$ variables is derived
from the time-evolution for $\rho$. Using operator identities,
this is equivalent to a zero-trace diffusion equation for the variables
$\bm{\alpha}$, of form: 
\begin{equation}
\dot{Q}\left(\bm{\lambda},t\right)=\mathcal{L}\left(\bm{\lambda}\right)Q\left(\bm{\lambda},t\right),\label{eq:q}
\end{equation}
where $\mathcal{L}\left(\bm{\lambda}\right)$ is the equivalent differential
operator for $Q$-function dynamics \citep{drummond-2020}.

We begin by utilizing Bohr's idea of measurement as amplification
to analyze a measurement $\hat{x}$ \citep{Bohr}. We consider a
single-mode field, using the simplest model of a measurement for
a field quadrature $\hat{x}$ $-$ that of direct amplification of
$\hat{x}$ -- modeled by the parametric Hamiltonian $H_{amp}=\frac{i\hbar g}{2}(\hat{a}^{\dagger2}-\hat{a}^{2})\,$
with $g>0$ \citep{yuen-1976}. In a rotating frame, we define complementary
quadrature phase amplitudes $\hat{x}=\hat{a}+\hat{a}^{\dagger}$ and
$\hat{p}=(\hat{a}-\hat{a}^{\dagger})/i$ where $\hat{a}$, $\hat{a}^{\dagger}$
are boson operators. We find  $\hat{x}\left(t\right)=\hat{x}\left(0\right)e^{gt}$
and $\hat{p}\left(t\right)=\hat{p}\left(0\right)e^{-gt}$.

We first solve for the measurement of $\hat{x}$ on a system in a
superposition $|\psi_{sup}\rangle$ of eigenstates $|x_{j}\rangle$
of $\hat{x}$. The $Q$ function for $|\psi_{sup}\rangle$ is a sum
of Gaussians, as well as sinusoidal terms, denoted by $\mathcal{I}$
\citep{milburne-holmes-supQ}. The $|x_{j}\rangle$ are realized as
highly squeezed states in $\hat{x}$, defined by $|x_{j},r\rangle_{sq}=D(x_{j}/2)S(r)|0\rangle$
where $r\rightarrow\infty$, $D(\beta_{0})=e^{\beta_{0}a^{\dagger}-\beta_{0}^{*}a}$
and $S(z)=e^{\frac{1}{2}(z^{*}a^{2}-za^{\dagger2})}$ \citep{yuen-1976},
implying $(\Delta\hat{x})^{2}=e^{-2r}$, $(\Delta\hat{p})^{2}=e^{2r}$
and $\langle\hat{x}\rangle=x_{j}$. We consider an initial superposition
at $t=t_{1}$ as $|\psi_{sup}\rangle=c_{1}|x_{1}\rangle+c_{2}e^{i\varphi}|x_{2}\rangle$,
and take $c_{1}$, $c_{2}$ real and $\varphi=\pi/2$. The $Q$-function
is
\begin{eqnarray}
Q(x,p,t_{1}) & = & \frac{e^{-p^{2}/2\sigma_{p}^{2}}}{2\pi\sigma_{x}\sigma_{p}}\Bigr(\sum_{j=1,2}|c_{j}|^{2}e^{-(x-x_{j})^{2}/2\sigma_{x}^{2}}\nonumber \\
 &  & -2c_{1}|c_{2}|e^{-[(x-x_{1})^{2}+(x-x_{2})^{2}]/4\sigma_{x}^{2}}\sin(px_{1}/\sigma_{x}^{2})\Bigl)\nonumber \\
\label{eq:Qsqsup}
\end{eqnarray}
with $\sigma_{x}^{2}=1+e^{-2r}$ and $\sigma_{p}^{2}=1+e^{2r}$.
For each pure eigenstate $|x_{j}\rangle$, the corresponding distribution
is a Gaussian with mean $x_{j}$ and variance $\sigma_{x}=1$ (as
$r\rightarrow\infty$). This indicates a ``hidden'' vacuum noise
level.

\begin{figure}
\begin{centering}
\includegraphics[width=0.9\columnwidth]{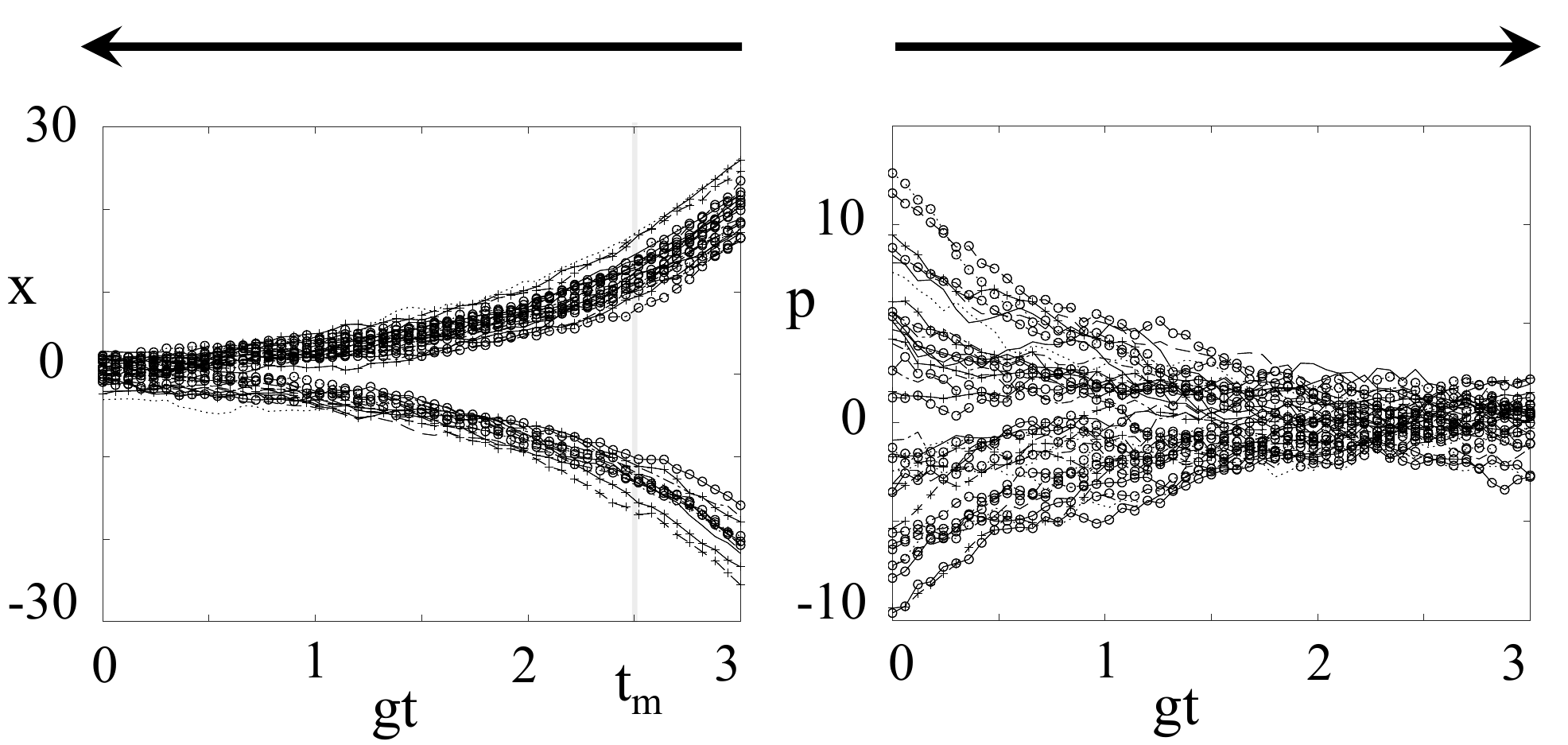}
\par\end{centering}
\caption{Solutions of (\ref{eq:backwardSDE}) and (\ref{eq:forwardSDE}) for
the measurement $\hat{x}$ on the system prepared in a superposition
$|\psi_{sup}\rangle$ of eigenstates of $\hat{x}$ (approximated by
(\ref{eq:Qsqsup})). The amplified $x$ propagates backwards, the
$\xi_{1}(t)$ sampled from (\ref{eq:margfx}). The attenuated $p$
propagates forwards. $x_{1}=-x_{2}=0.8$, $r=2$, $c_{1}=-ic_{2}=1/\sqrt{2}$.
\label{fig:x-trajectories-superposition-1}}
\end{figure}

\begin{figure}
\begin{centering}
\includegraphics[width=0.9\columnwidth]{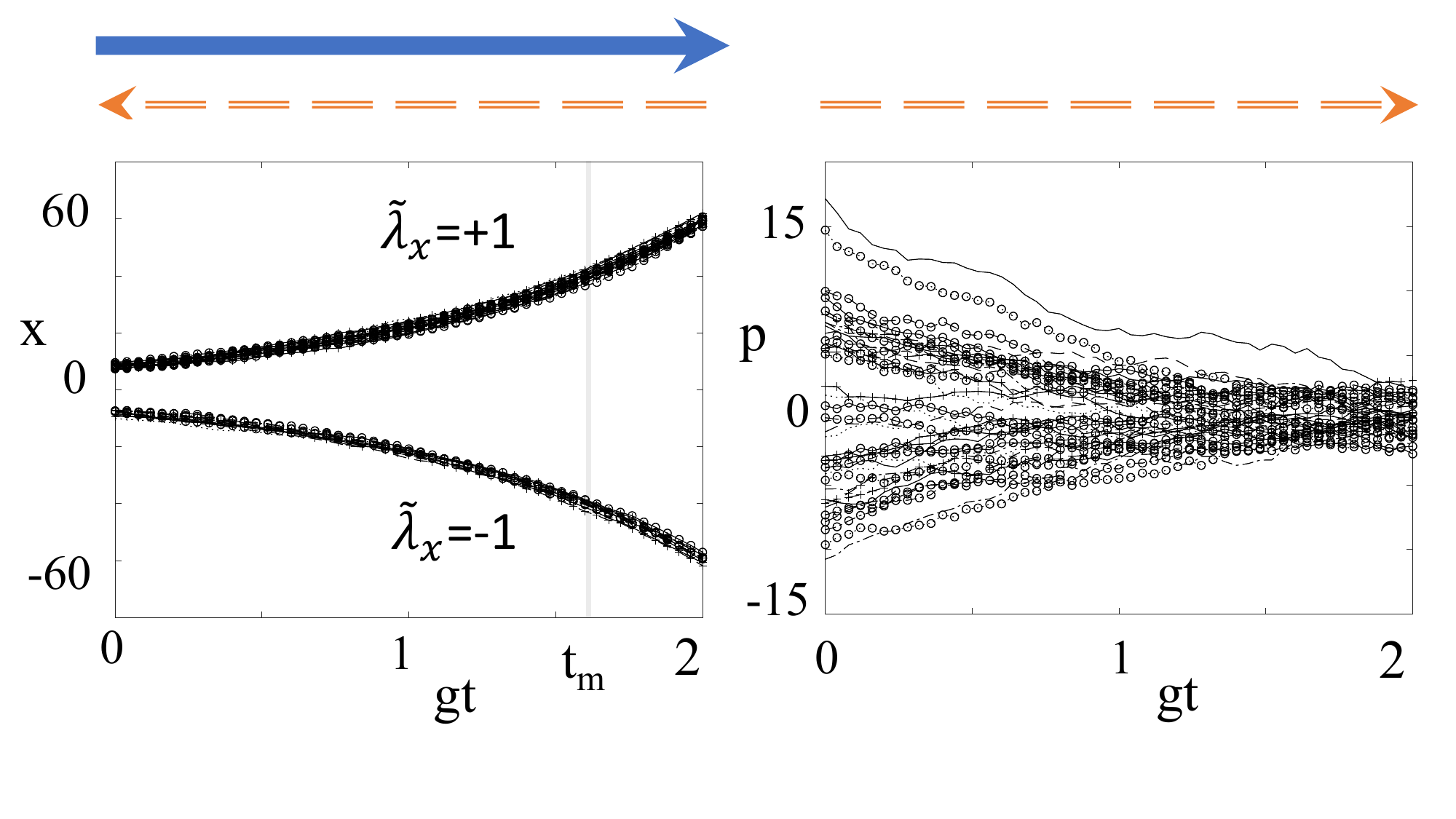}
\par\end{centering}
\begin{centering}
\includegraphics[width=0.8\columnwidth]{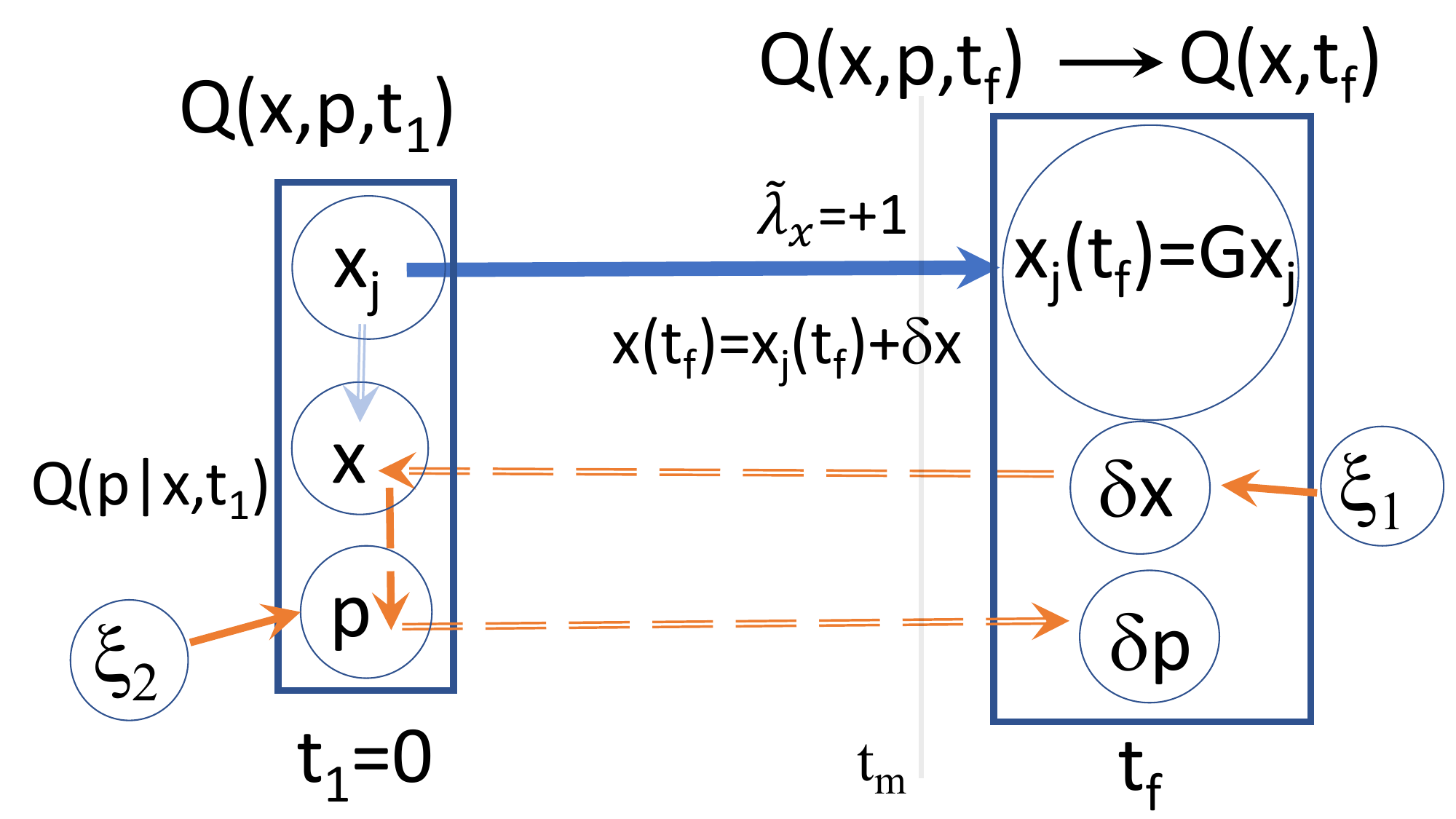}
\par\end{centering}
\begin{centering}
\includegraphics[width=1\columnwidth]{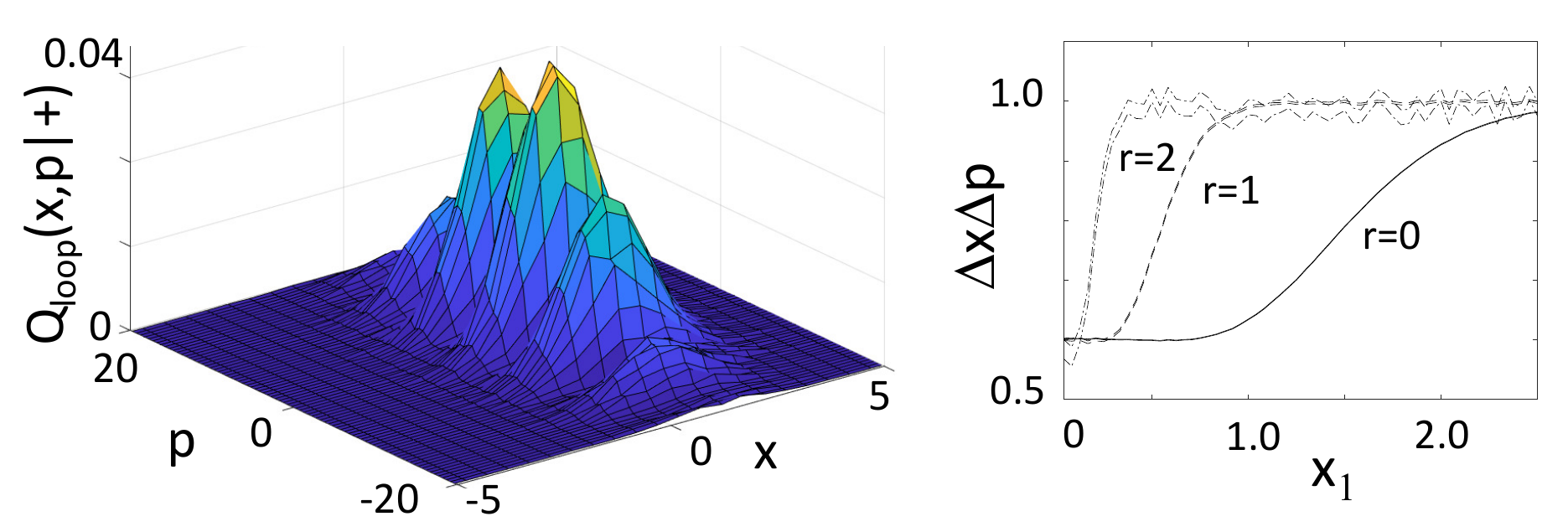}
\par\end{centering}
\caption{Top: Solutions of (3) and (4), for $|\psi_{sup}\rangle$ where $x_{1}=-x_{2}=8$,
$r=2$. Trajectories for $x$ (left), and  (right) trajectories
for $p$ connected to the positive branch $\widetilde{\lambda}_{x}=+1$.
Centre: Formation of a hidden loop. Lower: $Q_{loop}(x,p|+)$ corresponding
to the upper branch $\widetilde{\lambda}_{x}=+1$ (left) and (right)
the value of $\Delta\hat{x}\Delta\hat{p}$ for the ``state'' $Q_{loop}(x,p|+)$,
for various superpositions $|\psi_{sup}\rangle$. Here, $gt_{f}=4$.
The parallel lines give sampling-error bounds with $1.2\times10^{7}$
trajectories \citep{companion-paper}.\label{fig:x-trajectories-superposition-loop}}
\end{figure}

The equation (\ref{eq:q}) has equivalent phase-space trajectories
$\bm{\lambda}\left(t\right)$, which satisfy \textbf{neither} classical
dynamics \textbf{nor} the causal properties of the EPR-Bell local
realistic models. As proved in the companion paper, they satisfy a
time-symmetric yet stochastic dynamics \citep{companion-paper}.
To obtain a mathematically tractable equation given the traceless
noise matrix in $\mathcal{L}\left(\bm{\lambda}\right)$, a reversal
of $t$ is given for  $x$. The amplified variable $x$ satisfies
\begin{align}
\frac{dx}{dt_{-}} & =-gx+\xi_{1}(t_{-})\label{eq:backwardSDE}
\end{align}
which is solved backward in time from a future boundary condition
(BC) specified at the time $t_{f}$, after the interaction $H_{amp}$
has been completed. The complementary variable satisfies
\begin{equation}
\frac{dp}{dt}=-gp+\xi_{2}(t)\label{eq:forwardSDE}
\end{equation}
with a BC at the initial time $t_{1}=0$. The Gaussian random noises
$\xi_{\mu}\left(t\right)$ satisfy $\left\langle \xi_{\mu}\left(t\right)\xi_{\nu}\left(t'\right)\right\rangle =2g\delta_{\mu\nu}\delta\left(t-t'\right)$.
Thus, there is a forward-backward stochastic differential equation,
for individual trajectories. The trajectories for $x$ and $p$ decouple.
One propagates forward, one backwards in time (Figs. \ref{fig:x-trajectories-superposition-1}
and \ref{fig:x-trajectories-superposition-loop}).  

The future BC determines the sampling distribution for $x$ at $t_{f}$
and is given by the ``future'' marginal $Q(x,t_{f})=\int Q(x,p,t_{f})dp$,
but this is completely defined by the initial state (\ref{eq:Qsqsup}),
$H_{amp}$ and $t_{f}$:\textbf{\textcolor{black}{
\begin{eqnarray}
Q(x,t_{f}) & = & \frac{1}{\sqrt{2\pi}\sigma_{x}(t_{f})}\sum_{j}|c_{j}|^{2}e^{-[x-Gx_{j}]^{2}/2\sigma_{x}^{2}(t_{f})}.\label{eq:margfx}
\end{eqnarray}
}}\textcolor{black}{The $Q(x,t_{f})$ is a sum of Gaussian terms with
amplified means $Gx_{j}$, where $G=e^{gt_{f}}$. By contrast, the
variances are $\sigma_{x}^{2}(t)=1+G(\sigma_{x}^{2}-1)$, which for
$r\rightarrow\infty$ remain at the hidden vacuum level $\sigma_{x}=1$.
More generally for arbitrary $\varphi$, interference terms $\mathcal{I}$
appear but deamplif}y as $t_{f}\rightarrow\infty$.\textcolor{red}{{}
}\textcolor{black}{}\textcolor{red}{}\textcolor{black}{The marginal
(\ref{eq:margfx}) and $x$-trajectories become identical to those
of the mixture of eigenstates, $\rho_{mix}=\sum_{j}|c_{j}|^{2}|x_{j}\rangle\langle x_{j}|$,
for which $\mathcal{I}=0$.}

Causal relations can be deduced by tracking the amplitudes in the
simulation.\textcolor{black}{{} There is a mapping of the Gaussian with
mean $x_{j}$ in (\ref{eq:Qsqsup}) to a Gaussian with amplified mean
$Gx_{j}$ in (\ref{eq:margfx}). }While the eigenvalue $x_{j}$ is
amplified, the noise is not. Since $\xi_{1}$ is independent of $x$,
this is clarified by writing  $x(t)=x_{j}(t)+\delta x(t)$. The
$x_{j}(t)$ is the observable part of $x$, given by $\frac{dx_{j}}{dt}=-gx_{j},$
which has a causal, deterministic solution, $x_{j}(t)=x_{j}(0)e^{gt}$
(blue line in Fig. \ref{fig:x-trajectories-superposition-loop}).
The other part to $x(t)$ is not amplified and therefore not observed:
it is \emph{retrocausal, microscopic and stochastic}: $\frac{d(\delta x)}{dt_{-}}=-g(\delta x)+\xi_{1}(t_{-})$,
the starting value $\delta x(t_{f})$ being the deviation of $x(t_{f})$
from $x_{j}(t_{f})$. The solutions give a constant average noise
magnitude for $\delta x$ throughout the dynamics, leading to two
``branches'', denoted $\widetilde{\lambda}_{x}=\pm1$, as $t_{f}\rightarrow\infty$
(Fig. 2, top left). The trajectories for $p$ show a decay to the
hidden vacuum noise level $\delta p$, with $\sigma_{p}\rightarrow1$
as $t_{f}\rightarrow\infty$.

We now examine the origin of a hidden causal loop. There is a connection
between trajectories for $x$ and $p$, given by the conditional at
$t=t_{1}=0$, 
\begin{eqnarray}
Q(p|x,t_{1}) & = & \frac{e^{-p^{2}/2\sigma_{p}^{2}}}{\sigma_{p}\sqrt{2\pi}}\left\{ 1-\frac{\sin(2px_{1}/\sigma_{x}^{2})}{\cosh(2xx_{1}/\sigma_{x}^{2})}\right\} .\label{eq:cond}
\end{eqnarray}
For each backward trajectory, $x(t_{f})$ propagates to a value $x(t_{1})$
at $t_{1}=0$. A distribution of values $p(t_{1})$ for each $x(t_{1})$
is generated by $Q(p|x(t_{1}))$. Each such $p$ propagates forward
to a value $p(t_{f})$, giving a microscopic feedback. There is\emph{
no }such connection for $\rho_{mix}$, for which $Q(p|x(t_{1})=Q(p)$,
which establishes a link between a hidden loop and a superposition.
The distribution $Q_{loop}(x,p|+)$ defined as the joint density of
the connected trajectories for $x$ and $p$ \emph{conditioned} on
the \emph{positive} branch $\widetilde{\lambda}_{x}=1$ ($x(t_{f})>0$)
can be derived from (\ref{eq:cond}) (Fig. \ref{fig:x-trajectories-superposition-loop}).
This defines a ``state'' at the original time $t_{1}=0$, given
by a particular branch of the superposition. The distribution for
$p$ defined by $Q_{loop}(x,p|+)$ reveals fringes. This induces
a variance $\Delta\hat{p}$ \emph{below} that of the eigenstate $|x_{1}\rangle$
(modelled as $|x_{1},r\rangle_{sq}$), associated with the positive
branch. The uncertainties associated with future measurements of $\hat{x}$
or $\hat{p}$ on this ``state'' violate the Heisenberg relation
$\Delta\hat{x}\Delta\hat{p}\geq1$, justifying the terminology ``hidden
loop''. Causal consistency demands that the joint probability density
for $x$ and $p$ at time $t$ created from the forward and backward
amplitudes is given by $Q(x,p,t)$ \citep{deutsch-1991}, which is
a function of $t$ (not $t_{f}$). This is confirmed in \citep{companion-paper},
avoiding Garndfather paradoxes, and is possible because $\xi_{1}$
is independent of $t_{f}$.

The probability density of the amplitudes $x(t_{f})$ \emph{is} the
probability $P(x)=|\langle\psi_{sup}|x\rangle|^{2}$ for detecting
the value $x(t_{f})/G$, thus satisfying Born's rule, as $G\rightarrow\infty$.
This is because the eigenvalue $x_{j}$ amplifies to $Gx_{j}$, whereas
$\delta x$ and $\mathcal{I}$ do not amplify, and the Gaussian term
for $Gx_{j}$ of the marginal (\ref{eq:margfx}) is proportional to
$|c_{j}|^{2}$. Hence, an ontological model for measurement is proposed
\citep{drummond-2020}, where the trajectory values represent a realisation
of the system, the detected value being $x(t_{f})$ as $G\rightarrow\infty$.
The \emph{inferred} outcome for $\hat{x}$ is $\widetilde{x}(t_{f})=x(t_{f})/G$,
which as $G\rightarrow\infty$ is precisely one of the eigenvalues
$x_{j}$. Born's rule is explained without a ``collapse'', the
measurement dynamics being treated in identical fashion to any other
physical process.

The model shows consistency with macroscopic realism. The value $x(t_{f})$
predetermines the outcomes for $\hat{x}$. For macroscopic superpositions
$|\psi_{sup}\rangle$ ($|x_{1}-x_{2}|\gg1$), the trajectories are
a ``line'' giving a causal relation connecting $x_{j}$ at time
$t_{1}$ to $Gx_{j}$ at time $t_{f}$, with thickness $\delta x$
(the hidden vacuum noise level) (Fig. 2). For each trajectory, there
is a one-to-one correspondence between the branch of $x(t_{f})$ and
that of $x(t_{0})$. . In fact, any $|\psi_{sup}\rangle$ becomes
a macroscopic superposition, with sufficient amplification $G$ (Fig.
1 at $t_{m}$). At time $t_{m}$,  the scaled amplitude $\widetilde{x}(t_{m})$
 \emph{is} the outcome of the measurement $\hat{x}$. \textcolor{black}{
}In fact, the scaled amplitude $\widetilde{x}(t)$ can be regarded
as defining a \emph{band} of amplitudes of width $\delta x/G$, so
that $\widetilde{x}(t_{m})$ is more precise as $G\rightarrow\infty$\textcolor{black}{.
} In the model, the noise $\delta x$ in \textcolor{black}{$x(t_{m})$}
from the future BC does not impact the present reality\emph{, }defined
by $\widetilde{x}(t_{m})$, for $|\psi_{sup}(t_{m})\rangle$. Consider
the system in the amplified state $|\psi_{sup}(t_{m})\rangle=e^{-iH_{amp}t_{m}/\hbar}|\psi_{sup}\rangle$.
Prior to a measurement of $\hat{x}$, the system has a predetermined
outcome $\widetilde{\lambda}_{x}=\widetilde{x}(t_{m})$ for $\hat{x}$.
Recalling the definition in the introduction \citep{leggett-1985},
we see that MR emerges, with sufficient amplification $G$.

\begin{figure}
\begin{centering}
\includegraphics[width=0.85\columnwidth]{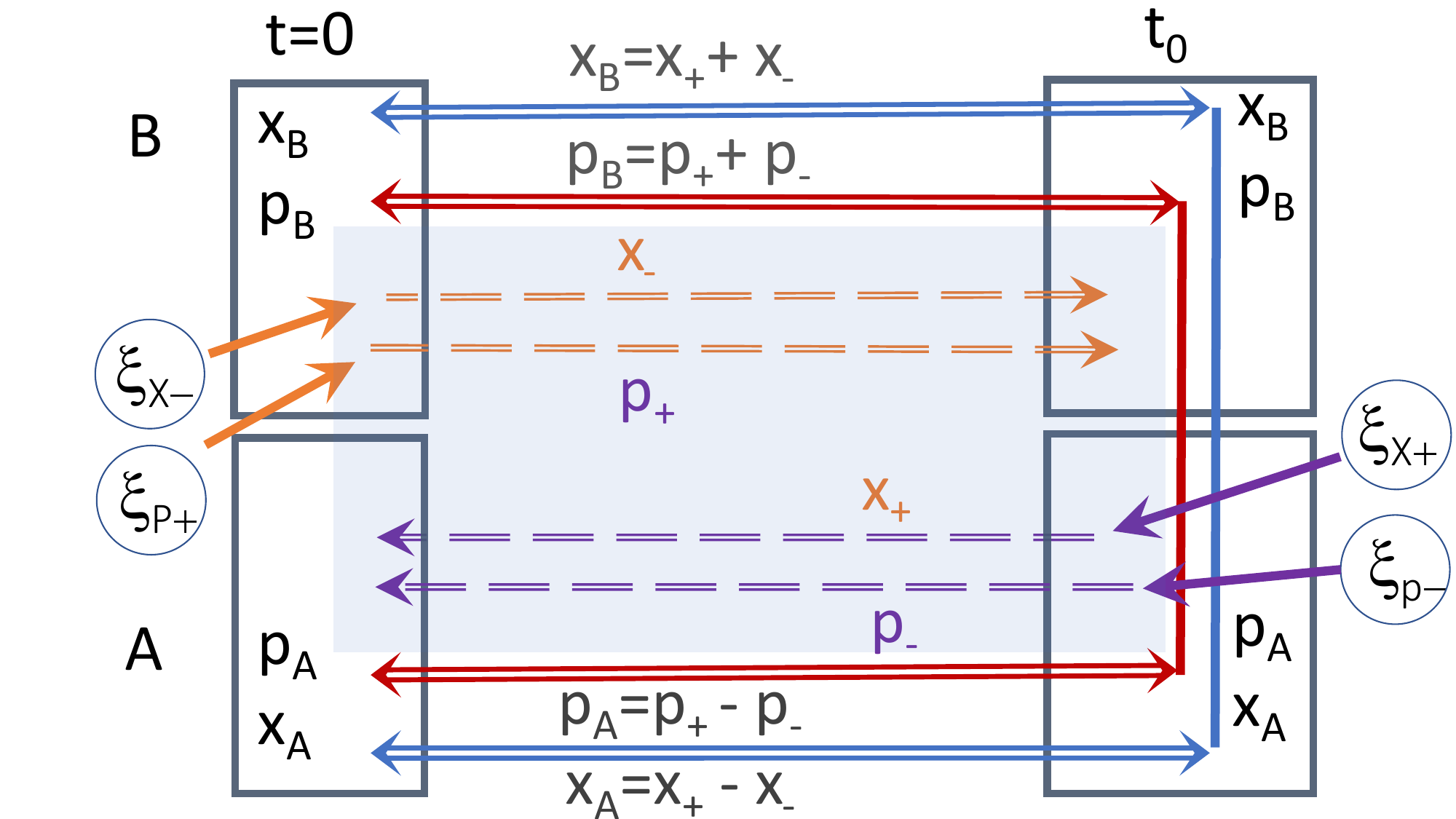}
\par\end{centering}
\caption{Creation of EPR entanglement for two modes $A$ and $B$ initially
in a vacuum state. The $x_{A}$, $x_{B}$, $p_{A}$ and $p_{B}$ are
combinations of forward and backward solutions $x_{\pm}$ and $p_{\pm}$.
Shared noise sources $\xi$ lead to correlations between $x_{A}$
and $x_{B}$ (and $p_{A}$ and $p_{B}$), depicted by the solid vertical
lines.\label{fig:entanglement}}
\end{figure}

A different type of forward-backward stochastic equations describes
the formation of entanglement (Fig. \ref{fig:entanglement}) \citep{sm}.
We solve for two field modes $A$ and $B$ that become entangled via
the interaction $H_{AB}=i\hbar\kappa(\hat{a}^{\dagger}\hat{b}^{\dagger}-\hat{a}\hat{b})$
\citep{reid-1989}. Boson operators $\hat{a}$, $\hat{b}$ and amplitudes
$\hat{x}_{A}$, $\hat{p}_{A}$, $\hat{x}_{B}$, $\hat{p}_{B}$ are
defined for each mode. After a time $t_{0}$, the system in a separable
vacuum state evolves to the EPR-entangled state $|\psi_{epr}\rangle$,
with $Q$ function $Q_{epr}(\bm{\lambda},t_{0})$ where $\bm{\lambda}=(x_{A},x_{B},p_{A},p_{B})$.
As $r=\kappa t_{0}\rightarrow\infty$, $|\psi_{epr}\rangle$ is an
eigenstate of $\hat{x}_{A}-\hat{x}_{B}$ and $\hat{p}_{A}+\hat{p}_{B}$
\citep{epr-1935}. The trajectories for amplitudes $x_{-}=x_{A}-x_{B}$
and $p_{+}=p_{A}+p_{B}$ are forward-propagating, and lead to squeezed
fluctuations. Those for $x_{+}=x_{A}+x_{B}$ and $p_{-}=p_{A}-p_{B}$
are backward-propagating and are amplified. The trajectories for
$x_{A}$ and $x_{B}$ (and $p_{A}$ and $p_{B}$) \emph{combine} the
forward and backward solutions, and become correlated \emph{as a result
of common noise sources} $\xi_{x\pm}$ (and $\xi_{p\pm}$).

\begin{figure}
\begin{centering}
\par\end{centering}
\begin{centering}
\includegraphics[width=0.85\columnwidth]{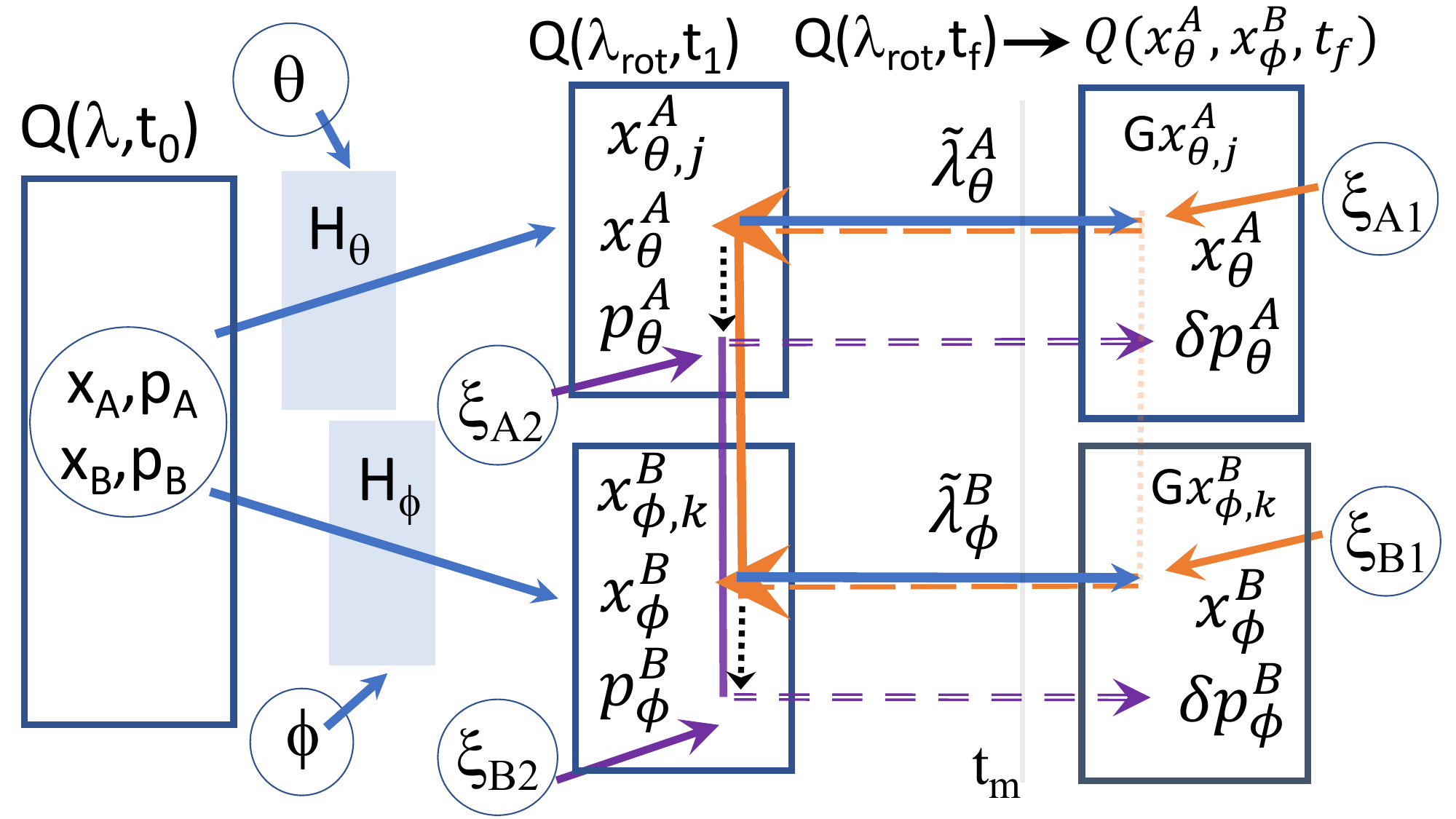}
\par\end{centering}
\begin{centering}
\par\end{centering}
\begin{centering}
\includegraphics[width=0.85\columnwidth]{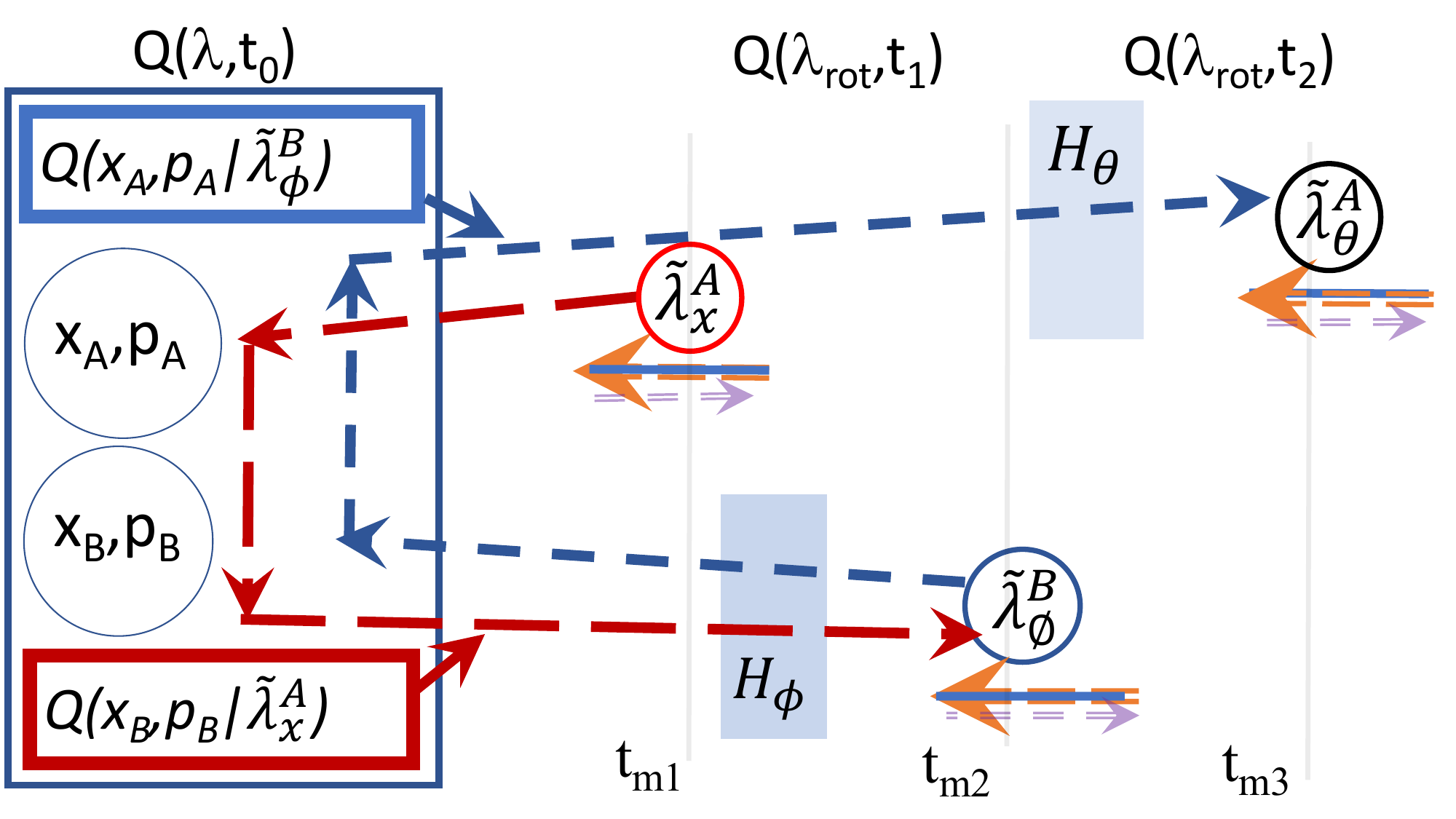}
\par\end{centering}
\caption{Top: Simulation of measurements $\hat{x}_{\theta}^{A}$ and $\hat{x}_{\phi}^{B}$
on EPR states.  The solid vertical lines at $t_{1}$ depict correlations
according to $Q(\bm{\lambda_{rot}},t_{1})$ which determine the measured
correlation at the boundary $t_{f}$ (dashed lines).  States violating
a Bell inequality show correlation between $x$ and $p$ giving local
hidden loops (dotted black lines). The amplified amplitudes $x_{\theta}^{A}(t_{m})$
and $x_{\phi}^{B}(t_{m})$ belong to one or other of the branches,
identified as outcomes $\widetilde{\lambda}_{\theta}^{A}$ and $\widetilde{\lambda}_{\phi}^{B}$.
Lower: Mechanism for nonlocality: The trajectories $x_{A}(t)$ corresponding
to a given branch $\widetilde{\lambda}_{x}^{A}$ trace back to time
$t_{0}$ where they only correlate with certain amplitudes of $B$,
hence restricting outcomes for a measurement $x_{\phi}^{B}$ (red
dashed line).\label{fig:entanglement-1} The restriction occurs regardless
of whether the operation $H_{\phi}$ has yet occurred at $B$. There
is no change to $\widetilde{\lambda}_{x}^{A}$ due to $H_{\phi}$,
although feedback from $H_{\phi}$ and $H_{amp}$ at $B$ (blue dashed
lines) allows Bell nonlocality if there is a further change of setting
($H_{\theta}$) at $A$.}
\end{figure}

A more complex set of equations describes a spin-$1/2$ system as
it becomes entangled with a meter. In the Supplementary Materials,
we derive forward-backward equations for combined system-meter amplitudes,
modeling the measurement of the spin of a system in a superposition
of eigenstates $|\pm\rangle$ of $\hat{\sigma}_{z}$ \citep{sm}.
 \textcolor{red}{}The spin outcome is determined by the final
meter amplitude $x_{c}$ which couples to  spin trajectories, so
that the distribution for the spin amplitudes conditioned on a $\pm1$
branch $\widetilde{\lambda}_{x_{c}}$ for $x_{c}$ is precisely the
$Q$ function of the eigenstate $|\pm\rangle$.

This motivates us to examine nonlocality. Figure \ref{fig:entanglement-1}
depicts measurement of EPR correlations using quadrature amplitude
measurements $\hat{x}_{\theta_{K}}^{K}=\hat{x}_{K}\cos\theta_{K}+\hat{p}_{K}\sin\theta_{K}$
where $K=A,B$. The $Q$ function for the initial state $|\psi\rangle$
is $Q(\bm{\lambda},t_{0})$. The choice of measurement settings
$\theta$ and $\phi$ is provided by phase-shift interactions $H_{\theta}^{A}$
and $H_{\phi}^{B}$ which give a local deterministic transformation
of the amplitudes: $x_{\theta}^{A}=x_{A}\cos\theta+p_{A}\sin\theta$,
$p_{\theta}^{A}=-x_{A}\sin\theta+p_{A}\cos\theta$ and similarly
at $B$. The evolved $Q$ function is $Q(\bm{\lambda_{rot}},t_{1})$
where $\bm{\lambda_{rot}}=(x_{\theta}^{A},p_{\theta}^{A},x_{\phi}^{B},p_{\phi}^{B})$.
The $\hat{x}_{\theta}^{A}$ and $\hat{x}_{\phi}^{B}$ are then measured
by an independent amplification $H_{amp}^{K}$ of $\hat{x}_{\theta_{K}}^{K}$
at each site $K$, as governed by forward-backward equations $\frac{dx_{\theta}^{K}}{dt_{-}}=-gx_{\theta_{K}}^{K}+\xi_{K1}\left(t\right)$
and $\frac{dp_{\theta_{K}}^{K}}{dt}=-gp_{\theta_{k}}^{K}+\xi_{K2}\left(t\right)$
where $\left\langle \xi_{K\mu}\left(t\right)\xi_{K'\nu}\left(t'\right)\right\rangle =2g\delta_{\mu\nu}\delta_{KK'}\delta\left(t-t'\right)$.
The boundary condition for the trajectories $x_{\theta}^{A}(t)$
and $x_{\phi}^{B}(t)$ is determined by the \emph{joint} \emph{marginal}
$Q(x_{\theta}^{A},x_{\phi}^{B},t_{f})$\textcolor{black}{. }The density
of the final amplitudes $x_{\theta}^{A}(t_{f})$ and $x_{\phi}^{B}(t_{f})$
coincides with the probabilities of outcomes for $|\psi\rangle$.
For some states, violation of a Bell inequality is possible \citep{gilchrist-prl-bell}.
A similar treatment models measurement of nonlocality using spin measurements
$\hat{\sigma}_{\theta}^{A}$ and $\hat{\sigma}_{\phi}^{B}$ \citep{sm}.

\begin{figure}
\begin{centering}
\includegraphics[width=1\columnwidth]{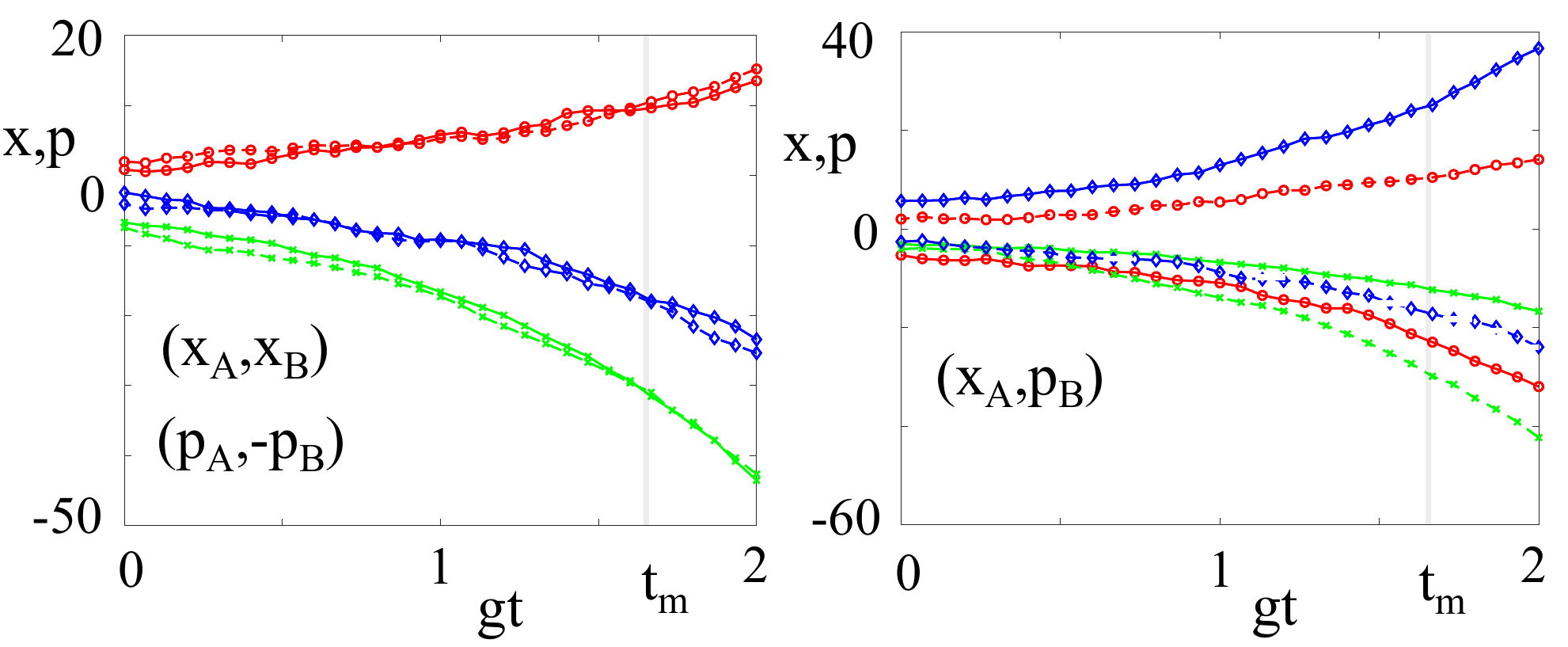}
\par\end{centering}
\caption{Simulation of the measurement of EPR correlations, revealing values
for $\widetilde{\lambda}_{\theta}^{A}$ and $\widetilde{\lambda}_{\phi}^{B}$.
Trajectories in the same run are shown with the same color/ symbol.
Left: Joint measurement of $\hat{x}_{A}$ and $\hat{x}_{B}$ on $|\psi_{epr}\rangle$,
$r=2$. The trajectories for $x_{A}$ and $x_{B}$ in the same run
are correlated. Trajectories $p_{A}$ and $-p_{B}$ for measurements
$\hat{p}_{A}$ and $-\hat{p}_{B}$ are given similarly. Right: Trajectories
$(x_{A},p_{B})$ for joint measurements of $\hat{x}_{A}$ and $\hat{p}_{B}$.
Here, $x_{A}$ ($p_{B}$) is given by the solid (dashed) line.\label{fig:epr}}
\end{figure}

The EPR correlations of $|\psi_{epr}\rangle$ are signified by measurements
of $\hat{x}_{K}$ or $\hat{p}_{K}$ at each site $K$ ($\theta,\phi=0$
or $\pi/2$). \textcolor{black}{ } Results for trajectories are
given in Fig.\textcolor{black}{{} \ref{fig:epr}}. \textcolor{black}{As
before, ``elements of reality'' consistent with macroscopic realism
emerge as the system amplifies.}\textcolor{black}{\emph{ }}\textcolor{black}{The
outcome inferred for $\hat{x}_{A}$ at time $t_{m}$ is $\widetilde{x}_{A}(t_{m})=x_{A}(t_{m})/G$,
which becomes sharp for large $G$. Hence, we identify a variable
$\widetilde{\lambda}_{x}^{A}=\widetilde{x}_{A}(t_{m})$ which predetermines
the outcome of a measurement $\hat{x}_{A}$ on the amplified state
$e^{-iH_{amp}^{A}t_{m}/\hbar}e^{-iH_{amp}^{B}t_{m}/\hbar}|\psi_{epr}\rangle$
at $t=t_{m}$. For large $r$, the correlation between $\widetilde{x}_{A}(t_{m})$
and $\widetilde{x}_{B}(t_{m})$ becomes perfect, so that the outcome
for $\hat{x}_{A}$ can be inferred from the measurement $\hat{x}_{B}$.
Similarly, after sufficient amplification at time $t_{m}$, the outcome
for $\hat{p}_{B}$ is given by $\widetilde{\lambda}_{p}^{B}=\widetilde{p}_{B}(t_{m})=p_{B}(t_{m})/G$.
For $r\rightarrow\infty$, there is perfect anticorrelation between
$\widetilde{p}_{A}(t)$ and $\widetilde{p}_{B}(t)$, so that the outcome
for $\hat{p}_{A}$ can be inferred from $\hat{p}_{B}$.}

\textcolor{black}{How is the prediction of $\hat{x}_{B}$ from $\hat{x}_{A}$
 (and similarly $\hat{p}_{B}$ from $\hat{p}_{A}$) possible? EPR's
reasoning, based on their premises, suggested that values  for $\hat{x}_{A}$
and $\hat{x}_{B}$ were predetermined and corrrelated at the initial
time $t_{0}$ $-$ but this assertion (for spins) was falsified
by Bell's theorem. Instead, we follow the procedure for a loop given
in Fig. 2 (Fig. \ref{fig:entanglement-1}, lower). The  trajectories
for $x_{A}$ }\textcolor{black}{\emph{conditioned}}\textcolor{black}{{}
on the value $\widetilde{\lambda}_{x}^{A}$ (which determines the
outcome for $\hat{x}_{A}$) can be traced }\textcolor{black}{\emph{back}}\textcolor{black}{{}
to the time $t_{0}$, where they correlate according to }$Q(\bm{\lambda},t_{0})$\textcolor{black}{{}
with certain amplitudes for $B$. This defines a distribution $Q_{loop}(x_{B},p_{B},p_{A}|\widetilde{\lambda}_{x}^{A})$,
which by integrating over the unobserved variable $p_{A}$, defines
$Q(x_{B},p_{B}|\widetilde{\lambda}_{x}^{A})$ for any measurement
at $B$. For $|\psi_{epr}\rangle,$ $Q(x_{B},p_{B}|\widetilde{\lambda}_{x}^{A})$
for $\widetilde{\lambda}_{x}^{A}=x_{A}$ corresponds to the $Q$ function
of the eigenstate $|x_{A}\rangle_{B}$ of $\hat{x}_{B}$.}

\textcolor{black}{A hidden loop is also provided by measurement of
$\hat{x}_{A}$ and $\hat{p}_{B}$ (Fig. 5, right). Such measurements
were considered by }Schrödinger \textcolor{black}{and }realized mesoscopically
by Colciaghi\textcolor{black}{{} et al \citep{schrodinger-1935}. }S\textcolor{black}{ince
$\hat{x}_{A}$ predicts $\hat{x}_{B}$}, Schrödinger\textcolor{black}{{}
asked whether the outcomes of $\hat{x}_{B}$ and $\hat{p}_{B}$ could
be simultaneously predetermined \citep{schrodinger-1935}? From the
simulation, at time $t_{m}$, such simultaneous values $\widetilde{\lambda}_{x}^{A}$,
$\widetilde{\lambda}_{p}^{B}$ can be identified by the pair of lines
in the same run. (NB: Fixing the setting at $A$ gives the future
BC that ensures $\widetilde{\lambda}_{x}^{A}$ }\textcolor{black}{\emph{is}}\textcolor{black}{{}
the value for $\hat{x}_{B}$, if the setting at $B$ is changed to
$\hat{x}_{B}$.)}

\textcolor{black}{Finally, despite aspects of retrocausality being
counterintuitive, we find consistency with a subset of local realistic
assumptions, given by three ``weak'' premises }\citep{weak,macro-bell}\textcolor{black}{.
Consider Figure \ref{fig:entanglement-1}. }\textbf{\textcolor{black}{\emph{Premise}}}\textcolor{black}{\emph{
}}\textbf{\textcolor{black}{\emph{(1)}}}\textbf{\textcolor{black}{:}}\textcolor{black}{{}
Consistency with }\textcolor{black}{\emph{macroscopic realism}}\textcolor{black}{{}
(MR). The variable $\widetilde{\lambda}_{\theta}^{A}$  at time $t_{m}$
predetermines the outcome }of the measurement $\hat{x}_{\theta}^{A}$
if made at $t_{f}$\textcolor{black}{. This value is not changed retrocausally.
}\textbf{\textcolor{black}{\emph{Premise (2):}}}\textcolor{black}{{}
}The value $\widetilde{\lambda}_{\theta}^{A}$ (with setting $\theta$
fixed) is not changed by any subsequent  change of setting $\phi$
at $B$, giving consistency with \emph{no-signaling}.  \textbf{\emph{Premise
(3):}} The predicted value \textcolor{black}{$\widetilde{\lambda}_{\theta,inf}^{B}$}
inferred from $A$ for the measurement $\hat{x}_{\theta}^{B}$ at
$B$ is a valid predetermination, \emph{provided the setting $\theta$
at $A$ is fixed}.

The subset  allows for EPR and Bell nonlocality. Examining Fig.
4 (lower), if the setting is changed to $\theta$ at $A$, then
the future BC, which is fixed by both settings, changes: Hence, while
$\widetilde{\lambda}_{\phi}^{B}$ is fixed, and the outcomes for $x_{\theta}^{A}$
are conditioned on $\widetilde{\lambda}_{\phi}^{B}$, there is no
longer the requirement that outcomes for \emph{further} measurements
$\phi'$ at $B$ are restricted by $Q(x_{B},p_{B}|\widetilde{\lambda}_{x}^{A})$:
An inferred prediction \textcolor{black}{$\widetilde{\lambda}_{x,inf}^{B}$
for $\hat{x}_{B}$ at $B$ is not a property for $B$ independent
of changes to the setting at $A$} (as was posited by EPR).  This
is understood because a change of setting at one location changes
only hidden terms in $Q$, but for changes of setting over both locations,
hidden terms (e.g. $\mathcal{I}$ in $Q(\bm{\lambda},t_{0})$) can
contribute to observable probabilities (e.g. in $Q(\bm{\lambda_{rot}},t_{2})$).
Examination of Bell's local-hidden-variables assumption reveals that,
for the model where the hidden variables correspond to the $\bm{\lambda}$
of $Q(\bm{\lambda},t_{0})$, Bell's locality assumption breaks down
when there are changes of settings at both locations. This adjustment
of settings means there is no conflict with the requirement of fine-tuning
for classical causal models \citep{wood-2015}. We also expect consistency
with other strict tests of quantum mechanics \citep{strict-tests,macro-bell}
which involve similar double rotations. As the interactions $H_{amp}$,
$H_{AB}$ and $H_{\theta}$ are realisable experimentally, and the
Q functions can be determined by tomography, experimental tests of
the forward-backward equivalences would seem feasible \citep{companion-paper}.
\begin{acknowledgments}
This research has been supported by the Australian Research Council
under Grants DP180102470 and DP190101480, and the Templeton Foundation
under Project Grant ID 62843. The authors thank NTT Research for their
financial and technical support.
\end{acknowledgments}

\begin{widetext} 

\section{Supplemental Materials: Generation of EPR entanglement}

Two separated modes $A$ and $B$ prepared in a two-mode squeezed
state 
\[
|\psi_{epr}\rangle=(1-\eta^{2})^{1/2}\sum_{n=0}^{\infty}\tanh^{n}r\thinspace|n\rangle_{A}|n\rangle_{B}
\]
at time $t_{0}$ possess EPR correlations. Here $\eta=\tanh r$ and
$|n\rangle_{A/B}$ are number states. Boson operators $\hat{a}$,
$\hat{b}$ and quadrature phase amplitudes $\hat{x}_{A}$, $\hat{p}_{A}$,
$\hat{x}_{B}$ and $\hat{p}_{B}$ are defined as in the main text
for each mode. The $Q$ function for $|\psi_{epr}\rangle$ is
\begin{eqnarray*}
Q_{epr}(\bm{\lambda}) & = & \frac{(1-\eta^{2})}{16\pi^{2}}e^{-\frac{1}{8}(x_{A}-x_{B})^{2}(1+\eta)}e^{-\frac{1}{8}(p_{A}+p_{B})^{2}(1+\eta)}e^{-\frac{1}{8}(x_{A}+x_{B})^{2}(1-\eta)}e^{-\frac{1}{8}(p_{A}-p_{B})^{2}(1-\eta)}
\end{eqnarray*}
where $\bm{\lambda}=(x_{A},x_{B},p_{A},p_{B})$. As $r\rightarrow\infty$,
$|\psi_{epr}\rangle$ is an eigenstate of $\hat{x}_{A}-\hat{x}_{B}$
and $\hat{p}_{A}+\hat{p}_{B}$. The EPR entangled state can be created
by the parametric down conversion process, for two-mode fields. The
Hamiltonian is 
\[
H_{AB}=i\hbar\kappa\left[\hat{a}^{\dagger}\hat{b}^{\dagger}-\hat{a}\hat{b}\right]
\]
where $\kappa$ is real.

We now choose for convenience to normalize the operators and amplitudes
differently from the main text, so that $\hat{X}_{A}=(\hat{a}+\hat{a}^{\dagger})/2$,
$\hat{P}_{A}=(\hat{a}-\hat{a}^{\dagger})/2i$ and $\hat{X}_{B}=(\hat{b}+\hat{b}^{\dagger})/2$,
$\hat{P}_{B}=(\hat{b}-\hat{b}^{\dagger})/2i$, implying $\alpha=x_{A}+ip_{A}$
and $\beta=x_{B}+ip_{B}$ for the Q function defined as $Q=\frac{1}{\pi^{2}}|\langle\alpha|\langle\beta|\psi_{r}\rangle|^{2}$.
This gives a variance product of $\Delta\hat{X_{A}}\Delta\hat{P}_{A}=1/4$
for a minimum uncertainty state. Here, $|\alpha\rangle|\beta\rangle$
is the two mode state with $|\alpha\rangle$ and $|\beta\rangle$
the coherent states for modes $A$ and $B$ respectively. We note
that the main text uses th different scaling given by $\alpha=(x_{A}+ip_{A})/2$
and $\beta=(x_{B}+ip_{B})/2$, which for a coherent state gives the
minimum uncertainty product of $\Delta\hat{X_{A}}\Delta\hat{P}_{A}=1$.

The initial state is the separable vacuum product state $|0\rangle|0\rangle$,
with Q function 
\[
Q\left(\bm{\lambda},0\right)=\frac{1}{\pi^{2}}e^{-(x_{A}^{2}+p_{A}^{2})}e^{-(x_{B}^{2}+p_{B}^{2})}
\]
The dynamics for the system evolving according to $H_{AB}$ is described
by a Fokker-Planck equation in the variables $\bm{\lambda}$. The
$x$ and $p$ variables decouple which greatly simplifies the solutions.
Transforming to $x_{\pm}=x_{A}\pm x_{B}$, $p_{\pm}=p_{A}\pm p_{B}$,
we obtain a Fokker-Planck equation with positive and negative diffusion.
The noise is restricted to second-order derivatives, and path-integral
theorems allow transformation to stochastic equations as before. We
find
\[
\frac{d}{dt_{-}}x_{+}=-\kappa x_{+}+\xi'_{3},\frac{d}{dt_{-}}p_{-}=-\kappa p_{-}+\xi'_{4}
\]
with boundary conditions in the future, and
\[
\frac{d}{dt}x_{-}=-\kappa x_{-}+\xi_{1},\frac{d}{dt}p_{+}=-\kappa p_{+}+\xi_{2}
\]
with boundary conditions in the past, where $t_{-}=-t$. The Gaussian
random noises $\xi_{\mu}\left(t\right)$ satisfy $\left\langle \xi_{\mu}\left(t\right)\xi_{\nu}\left(t'\right)\right\rangle =4\kappa\delta_{\mu\nu}\delta\left(t-t'\right)$.

The final Q function after evolution for a time $t_{0}$ is given
as
\begin{eqnarray*}
Q_{epr}(\bm{\lambda},t_{0}) & = & \frac{(1-\tanh^{2}r)}{\pi}e^{-\frac{1}{2}(x_{A}-x_{B})^{2}(1+\tanh r)}e^{-\frac{1}{2}(p_{A}+p_{B})^{2}(1+\tanh r)}e^{-\frac{1}{2}(x_{A}+x_{B})^{2}(1-\tanh r)}e^{-\frac{1}{2}(p_{A}-p_{B})^{2}(1-\tanh r)}
\end{eqnarray*}
 with $r=\kappa t$. In the new coordinates, this is 
\begin{eqnarray*}
Q_{epr}(\bm{\lambda},t_{f}) & = & \frac{(1-\eta^{2})}{\pi^{2}}e^{-\frac{1}{2}x_{-}^{2}(1+\eta)}e^{-\frac{1}{2}p_{+}^{2}(1+\eta)}\times e^{-\frac{1}{2}x_{+}^{2}(1-\eta)}e^{-\frac{1}{2}p_{-}^{2}(1-\eta)}
\end{eqnarray*}
where $\eta=\tanh r$. In order to carry out the simulation, the final
marginal at the time $t_{0}$ is evaluated for each variable from
the final $Q$ function. In each case, we see that the marginal, which
provides the boundary condition for the dynamics, is a simple Gaussian.
The boundary condition for the simulation is simpler than the case
given by Figures 1 and 2 in the main text, and follows the same procedure
as given above in the companion paper  \citep{companion-paper},
for the measurement of the entanglement of the EPR state $|\psi_{epr}\rangle$.

The ``position'' difference and ``momentum'' sum become squeezed,
giving the EPR correlations, the dynamics for these variables being
in the forward direction. The final variances as $r\rightarrow\infty$
($\eta\rightarrow1$) are at the level of $1/2$ (which corresponds
to $2$ using the normalization of the amplitudes as in the main text),
which corresponds to the eigenstate of the sums and difference, as
expected for the EPR correlations. The ``position'' sum and ``momentum''
difference are amplified. The individual trajectory values for $x_{A}$,
$p_{A}$, $x_{B}$ and $p_{B}$ are determined by adding and subtracting
the solutions for $x_{\pm}$, $p_{\pm}$.

\section{Supplemental Materials: Measurement of Spin}

\subsection{Spin measurement model}

Here, we give solutions for the measurement of the spin $\sigma_{z}$
of a spin -$1/2$ system $A$. The spin $1/2$ (qubit) system is coupled
to a meter, modelled as a field mode $C$ prepared initially in a
coherent state $|\gamma\rangle_{C}$. The measurement is made by coupling
the meter and spin-$1/2$ system via the Hamiltonian 
\begin{eqnarray*}
H_{I} & = & \hbar\chi\sigma_{z}n_{c}=\hbar\chi(a_{+}^{\dagger}a_{+}-a_{-}^{\dagger}a_{-})n_{c}
\end{eqnarray*}
The $\sigma_{z}$ is the Pauli spin operator for the qubit system
$A$, $n_{c}$ is the number operator for the field mode $C$, and
$G$ is a real constant. We may consider that the spin $1/2$ system
comprises two modes denoted $A_{+}$ and $A_{-}$ respectively, with
respective boson operators $a_{+}$ and $a_{-}$, so that $\sigma_{z}=a_{+}^{\dagger}a_{+}-a_{-}^{\dagger}a_{-}$,
$\sigma_{x}=a_{+}^{\dagger}a_{-}-a_{-}^{\dagger}a_{+}$ and $\sigma_{y}=(a_{+}^{\dagger}a_{-}-a_{-}^{\dagger}a_{+})/i$.
The up and down eigenstates of $\sigma_{z}$ are $|\uparrow\rangle=|1\rangle_{+}|0\rangle_{-}$
and $|\downarrow\rangle=|0\rangle_{-}|1\rangle_{+}$, where $|n\rangle_{\pm}$
denotes an eigenstate of $a_{\pm}^{\dagger}a_{\pm}$ for mode $A_{\pm}$.

Let us consider the general qubit superposition state 
\[
|\psi_{sup,s}\rangle=c_{1}|\uparrow\rangle+c_{2}|\downarrow\rangle=c_{1}|1\rangle_{+}|0\rangle_{-}+c_{2}|0\rangle_{+}|1\rangle_{-}
\]
where $c_{1}$ and $c_{2}$ are complex amplitudes. After an interaction
time $T$ such that $\chi T=\pi/2$, the evolution under $H_{I}$
gives the final entangled state 
\begin{eqnarray*}
|\psi_{ent}\rangle & = & c_{1}|\uparrow\rangle|\gamma_{0}\rangle+c_{2}|\downarrow\rangle|-\gamma_{0}\rangle\\
 & = & c_{1}|1\rangle_{a+}|0\rangle_{a-}|\gamma_{0}\rangle+c_{2}|0\rangle_{a+}|1\rangle_{a-}|-\gamma_{0}\rangle
\end{eqnarray*}
where here we have selected for convenience the initial phase of
the meter to be complex, so that we write $\gamma=i\gamma_{0}$ where
$\gamma_{0}$ is real. 

We note that the quadrature phase amplitude measurement $\hat{X}_{c}$
on the meter mode will indicate the outcome for the spin measurement:
The sign of the outcome for $\hat{X}_{c}$ if positive gives the ``up''
outcome and if negative gives the ``down'' outcome for the spin
$1/2$ measurement. The final amplification can be modelled via the
solutions given by $H_{amp}$ in the main text. However, the measurement
interaction $H_{I}$ has entangled the meter with the spin $1/2$
system. 

\subsection{Fokker-Planck equation of the Q function}

The equation of motion for the three-mode Q function can be derived
from the equation of motion for the density operator $\rho$ using
standard methods. We denote the Q function by $Q(\alpha)$ where the
amplitudes for the modes $C$, $A_{+}$ and $A_{-}$ by $\alpha_{c}$,
$\alpha_{+}$ and $\alpha_{-}$ respectively, so that $\alpha\equiv(\alpha_{c},\alpha_{+},\alpha_{-})$.
 We use the well-known correspondences
\[
\hat{a}\rho\rightarrow(\alpha+\frac{\partial}{\partial\alpha^{*}})Q(\alpha)
\]
\[
\rho\hat{a}\rightarrow\alpha Q(\alpha)
\]
\[
\rho\hat{a}^{\dagger}\rightarrow(\alpha^{*}+\frac{\partial}{\partial\alpha})Q(\alpha)
\]
\[
\hat{a}^{\dagger}\rho\rightarrow\alpha^{*}Q(\alpha)
\]
to deduce the Fokker-Planck (FP) equation
\begin{eqnarray}
\frac{dQ}{dt} & = & i\chi\frac{\partial}{\partial\alpha_{c}}\{\alpha_{c}(|\alpha_{+}|^{2}-|\alpha_{-}|^{2})Q\}+i\chi\frac{\partial}{\partial\alpha_{+}}\{\alpha_{+}(|\alpha_{c}|^{2}-1)Q\}\nonumber \\
 &  & -i\chi\frac{\partial}{\partial\alpha_{-}}\{\alpha_{-}(|\alpha_{c}|^{2}-1)Q\}+"cc"+i\chi\frac{\partial^{2}\alpha_{c}\alpha_{+}Q}{\partial\alpha_{c}\partial\alpha_{+}}-i\chi\frac{\partial^{2}\alpha_{c}\alpha_{-}Q}{\partial\alpha_{c}\partial\alpha_{-}}\label{eq:fp}
\end{eqnarray}
where $"cc"$ denotes the complex conjugate of the previous terms.

We are interested in the limit of a very large meter field $C$, where
$\gamma_{0}$ is very large. Here, the initial values of $\alpha_{c}$
are centred around $\gamma_{0}$. Hence we define the variables $\alpha_{\pm}$,
$\widetilde{\alpha}_{c}=\frac{\alpha_{c}}{\gamma_{0}}$ , and rewrite
the FP equation as 
\begin{eqnarray*}
\frac{dQ}{d(\chi t)} & = & i\frac{1}{|\gamma_{0}|^{2}}\frac{\partial}{\partial\widetilde{\alpha}_{c}}\{\widetilde{\alpha}_{c}(|\alpha_{+}|^{2}-|\alpha_{-}|^{2})Q\}+i\frac{\partial}{\partial\alpha_{+}}\{\alpha_{+}(|\widetilde{\alpha}_{c}|^{2}-\frac{1}{|\gamma_{0}|^{2}})Q\}\\
 &  & -i\frac{\partial}{\partial\alpha_{-}}\{\alpha_{-}(|\widetilde{\alpha}_{c}|^{2}-\frac{1}{|\gamma_{0}|^{2}})Q\}+"cc"+i\frac{1}{|\gamma_{0}|^{2}}\frac{\partial^{2}\tilde{\alpha}_{c}\alpha_{+}Q}{\partial\widetilde{\alpha}_{c}\partial\alpha_{+}}-i\frac{1}{|\gamma_{0}|^{2}}\frac{\partial^{2}\widetilde{\alpha}_{c}\alpha_{-}Q}{\partial\widetilde{\alpha}_{c}\partial\alpha_{-}}
\end{eqnarray*}
where $\frac{1}{|\gamma_{0}|^{2}}$ is small. Solving for the dominant
drift terms, ignoring noise at first, we find
\begin{eqnarray*}
\dot{\alpha}_{+} & = & -i\chi\alpha_{+}(|\alpha_{c}|^{2}-1)\\
\dot{\alpha_{-}} & = & i\chi\alpha_{-}(|\alpha_{c}|^{2}-1)
\end{eqnarray*}
Hence on defining $n_{+}=\alpha_{+}^{*}\alpha_{+}$ and $n_{-}=\alpha_{-}^{*}\alpha_{-}$,
we see that
\begin{eqnarray*}
\dot{n}_{+} & = & \dot{\alpha_{+}}\alpha_{+}^{*}+\alpha_{+}\dot{\alpha_{+}^{*}}=-i\chi\alpha_{+}\alpha_{+}^{*}(|\alpha_{c}|^{2}-1)+cc=0
\end{eqnarray*}
and similarly, $\dot{n}_{-}=0$. Introducing $n_{spin}=n_{+}-n_{-}$,
we see that $\dot{n}_{spin}=0$. Hence, amplitudes $n_{\pm}$ and
$n_{c}=\alpha_{c}^{*}\alpha_{c}$, and also the spin term $n_{spin}=n_{+}-n_{-}$,
will be constant in this limit.

We next examine the feedback to the meter mode, at first ignoring
the feedback from the second-derivative terms which are in fact of
similar order to the deterministic feedback. The simplified feedback
due to the deterministic terms is given by
\[
\dot{\alpha_{c}}=-i\chi\alpha_{c}(n_{spin})
\]
The solution in this approximation is ($\chi T=\pi/2$)
\begin{eqnarray*}
\alpha_{c} & = & \alpha_{c}(0)e^{-i\chi n_{spin}T}=\alpha_{c}(0)e^{-in_{spin}\pi/2}=-i\alpha_{c}(0)\sin(n_{spin}\pi/2)
\end{eqnarray*}
Now, $\alpha_{c}(0)\sim i\gamma_{0}$, so that $\alpha_{c}(T)\sim\gamma_{0}\sin(n_{spin}\pi/2)$.
We see that where the spin of system $A$ is $n_{spin}\sim1$, the
final value of the meter amplitude will be $\alpha_{c}(T)\sim\gamma_{0}$,
while where the spin of system $A$ is $n_{spin}\sim-1$, the final
value of the meter amplitude is $\alpha_{c}(T)\sim-\gamma_{0}$. This
simple analysis of the feedback on the meter is not sufficient to
fully explain the system-meter coupling, which requires the second
derivative (noise terms) to be included for a complete treatment.
This is because the distribution of the initial values of $n_{spin}$
is given by the amplitudes of the Q function

\begin{eqnarray*}
Q_{initial} & = & =\frac{1}{\pi^{2}}|\langle\alpha_{+}|\langle\alpha_{-}|\psi_{sup,s}\rangle|^{2}=\frac{e^{-|\alpha_{+}|^{2}-|\alpha_{-}|^{2}}}{\pi^{2}}|c_{1}\alpha_{+}^{*}+c_{2}\alpha_{-}^{*}|^{2}
\end{eqnarray*}
which are \emph{not} all centred at either $+1$ or $-1$. The function
reduces to (taking $c_{1}=c_{2}$)
\begin{eqnarray*}
Q_{initial} & \rightarrow & \frac{e^{-(x_{A+}^{2}+p_{A+}^{2}+x_{A-}^{2}+p_{A-}^{2})}}{2\pi^{2}}\{(x_{A+}^{2}+p_{A+}^{2})+(x_{A-}^{2}+p_{A-}^{2})+x_{A+}x_{A-}+p_{A+}p_{A-}\}
\end{eqnarray*}
where we have expressed the amplitudes in terms of real and imaginary
parts
\begin{eqnarray*}
\alpha_{+} & = & x_{A+}+ip_{A+}\\
\alpha_{-} & = & x_{A-}+ip_{A-}
\end{eqnarray*}
Here, $n_{spin}=x_{A+}^{2}+p_{A+}^{2}-x_{A-}^{2}-p_{A-}^{2}$. 

This point can be further made by considering the system prepared
in the eigenstate $|\uparrow\rangle$, for which the final state of
the meter according to the quantum Hamiltonian $H_{I}$ is always
$|\gamma_{0}\rangle$, implying that the final value of the meter-amplitude
$\alpha_{c}(T)$ is always near $\gamma_{0}$. On transforming to
polar coordinates, and then to coordinates $n_{\pm}$, the $Q$ function
for the eigenstate $|\uparrow\rangle$ is written\textcolor{black}{
\[
Q_{\uparrow}(n_{+},n_{-})=e^{-n_{+}-n_{-}}n_{+}
\]
}where $n_{+}=x_{A+}^{2}+p_{A+}^{2}$ and $n_{-}=x_{A-}^{2}+p_{A-}^{2}$.
\textcolor{black}{Defining $n_{spin}=n_{+}-n_{-}$ }and $n_{sum}=n_{+}+n_{-}$,\textcolor{black}{{}
we rewrite this as }
\[
Q_{\uparrow}(n_{sum},n_{spin})=e^{-n_{sum}}(n_{sum}+n_{spin})/4
\]
\textcolor{black}{where we note that $n_{sum}$ is restricted such
that $n_{sum}\geq|n_{spin}|$. }The marginal distribution for $n_{spin}$
is hence
\[
P(n_{spin})=\frac{e^{-|n_{spin}|}}{4}\{n_{spin}+|n_{spin}|+1\}
\]
Hence, using the simple treatment which ignores diffusion terms,
we see that there is a probability of $1/4$ that the initial value
$n_{spin}$ is negative, which implies (according to the simple treatment)
a significant probability that the final amplitude $\alpha_{c}(T)$
(given by $\alpha_{c}(T)\sim\gamma_{0}\sin(n_{spin}\pi/2)$ in the
simple treatment) is significantly different to $\gamma_{0}$. The
noise terms are hence necessary to fully account for the properties
of the meter.

\subsubsection{Transforming coordinates}

In order to properly investigate the noise terms in the FP equation,
FP equation (\ref{eq:fp}) is further transformed. We take logarithmic
variables, as is used in the similar treatment of the anharmonic oscillator.
If $\alpha_{j}=e^{\theta_{j}}$, and we define a complex phase $\phi_{j}=\theta_{j}/\sqrt{i}$
then: 
\begin{eqnarray*}
\alpha_{j} & = & e^{\sqrt{i}\phi_{j}}\\
n_{j} & = & \alpha_{j}\alpha_{j}^{*}
\end{eqnarray*}
where $j\in\{+,-,c\}$ denotes modes $A_{+}$, $A_{-}$ and $C$,
also symbolised by $a_{+}$, $a_{-}$ and $c$. Here, we define for
mode $A$ the difference $n_{spin}=n_{+}-n_{-}$.

We proceed to demonstrate the result of changing variables in the
distribution, as follows. In $\phi$ coordinates the distribution
is modified by the Jacobian of the transformation: 
\begin{eqnarray*}
Q_{\phi} & = & \left\langle e^{\sqrt{i}\phi}\left|\hat{\rho}\right|e^{\sqrt{i}\phi}\right\rangle \left|\frac{\partial\alpha}{\partial\phi}\right|=Q\left|\frac{\partial\alpha}{\partial\phi}\right|=Q\alpha\alpha^{*}
\end{eqnarray*}
There is also a chain rule for derivatives when changing variables,
\[
\frac{\partial}{\partial\alpha}=\frac{\partial\theta}{\partial\alpha}\frac{\partial}{\partial\theta}=\frac{1}{\sqrt{i}\alpha}\frac{\partial}{\partial\theta}
\]
To transform to phase coordinates, we first transform the Fokker-Planck
equation into a re-ordered form: The effect of the Jacobian is
such that it does not alter the equation for $Q_{\phi}$, apart from
the variable change :
\begin{eqnarray*}
\frac{\partial Q}{\partial t} & = & \Biggl[\mbox{i}\chi\alpha_{c}\frac{\partial}{\partial\alpha_{c}}\left[\alpha_{+}\alpha_{+}^{*}-\alpha_{-}\alpha_{-}^{*}\right]+i\chi\alpha_{+}\frac{\partial}{\partial\alpha_{+}}\{|\alpha_{c}|^{2}\}-i\chi\alpha_{-}\frac{\partial}{\partial\alpha_{-}}\{|\alpha_{c}|^{2}\}\\
 &  & +i\chi\left[\alpha_{c}\frac{\partial}{\partial\alpha_{c}}\right]\{\alpha_{+}\frac{\partial}{\partial\alpha_{+}}\}-i\chi\left[\alpha_{c}\frac{\partial}{\partial\alpha_{c}}\right]\{\alpha_{-}\frac{\partial}{\partial\alpha_{-}}\}+cc\Biggl]Q
\end{eqnarray*}
We get
\begin{eqnarray*}
\frac{\partial Q_{\phi}}{\partial t} & = & \chi\{\sqrt{i}\frac{\partial}{\partial\phi_{c}}\left(n_{+}-n_{-}\right)+\sqrt{i}\frac{\partial}{\partial\phi_{+}}\left(n_{c}\right)-\sqrt{i}\frac{\partial}{\partial\phi_{-}}\left(n_{c}\right)\\
 &  & +\frac{\partial}{\partial\phi_{c}}\frac{\partial}{\partial\phi_{+}}-\frac{\partial}{\partial\phi_{c}}\frac{\partial}{\partial\phi_{-}}+"cc"\}Q
\end{eqnarray*}
Continuing we let $\phi_{j}=\left(z_{j}+iy_{j}\right)/\sqrt{2}$
\[
\frac{\partial}{\partial\phi}=\frac{1}{\sqrt{2}}\left[\frac{\partial}{\partial z}+\frac{\partial}{i\partial y}\right]
\]
Since $\sqrt{i}=(1+i)/\sqrt{2}$, we have
\begin{align*}
\alpha & =e^{\left(1+i\right)\left(z+iy\right)/2}\\
n & =e^{z-y}
\end{align*}
We find
\begin{eqnarray*}
\sqrt{i}\frac{\partial}{\partial\phi}+\sqrt{-i}\frac{\partial}{\partial\phi^{*}} & = & \frac{(1+i)}{2}\left[\frac{\partial}{\partial z}+\frac{\partial}{i\partial y}\right]+\frac{(1-i)}{2}\left[\frac{\partial}{\partial z}-\frac{\partial}{i\partial y}\right]\\
 & = & \frac{\partial}{\partial z}+\frac{\partial}{\partial y}
\end{eqnarray*}
\begin{eqnarray*}
\frac{\partial^{2}}{\partial\phi^{2}}+\frac{\partial^{2}}{\partial\phi^{*2}} & = & \frac{1}{2}\left[\frac{\partial}{\partial z}+\frac{\partial}{i\partial y}\right]^{2}+\frac{1}{2}\left[\frac{\partial}{\partial z}-\frac{\partial}{i\partial y}\right]^{2}\\
 & = & \frac{\partial^{2}}{\partial z^{2}}-\frac{\partial^{2}}{\partial y^{2}}
\end{eqnarray*}
\begin{eqnarray*}
\frac{\partial^{2}}{\partial\phi_{c}\partial\phi_{+}}+\frac{\partial^{2}}{\partial\phi_{c}^{*}\partial\phi_{+}} & = & \frac{1}{2}\left[\frac{\partial}{\partial z_{c}}+\frac{\partial}{i\partial y_{c}}\right]\left[\frac{\partial}{\partial z_{+}}+\frac{\partial}{i\partial y_{+}}\right]+\frac{1}{2}\left[\frac{\partial}{\partial z_{c}}-\frac{\partial}{i\partial y_{c}}\right]\left[\frac{\partial}{\partial z_{+}}-\frac{\partial}{i\partial y_{+}}\right]\\
 & = & \frac{\partial^{2}}{\partial z_{x}\partial z_{+}}-\frac{\partial^{2}}{\partial y_{c}\partial y_{+}}
\end{eqnarray*}
Hence our final equation is
\begin{eqnarray*}
\frac{\partial Q}{\partial t} & = & \chi\{\frac{\partial}{\partial z_{c}}\left(n_{spin}\right)+\frac{\partial}{\partial y_{c}}\left(n_{spin}\right)\\
 &  & +\frac{\partial}{\partial z_{+}}\left(n_{c}\right)+\frac{\partial}{\partial y_{+}}\left(n_{c}\right)-\frac{\partial}{\partial z_{-}}\left(n_{c}\right)-\frac{\partial}{\partial y_{-}}\left(n_{c}\right)\\
 &  & +\frac{\partial}{\partial z_{c}}\frac{\partial}{\partial z_{+}}-\frac{\partial}{\partial y_{c}}\frac{\partial}{\partial y_{+}}-\frac{\partial}{\partial z_{c}}\frac{\partial}{\partial x_{-}}+\frac{\partial}{\partial y_{c}}\frac{\partial}{\partial y_{-}}\}Q
\end{eqnarray*}
where $n_{spin}=n_{+}-n_{-}$ represents the ``spin'' of system
$A$. 

\subsubsection*{Further transformations}

We now transform to new variables.  We define for the system $A$
the new variables 
\begin{eqnarray*}
z_{++} & = & \left(z_{+}+z_{-}\right)/2\\
z_{+-} & = & \left(z_{+}-z_{-}\right)/2
\end{eqnarray*}
 and 
\begin{eqnarray*}
y_{++} & = & \left(y_{+}+y_{-}\right)/2\\
y_{+-} & = & \left(y_{+}-y_{-}\right)/2
\end{eqnarray*}
 We then define the coupled meter-system variables 
\begin{eqnarray*}
z_{c+} & = & \left(z_{c}+z_{+-}\right)\\
z_{c-} & = & \left(z_{c}-z_{+-}\right)
\end{eqnarray*}
and 
\begin{eqnarray*}
y_{c+} & = & \left(y_{c}+y_{+-}\right)\\
y_{c-} & = & \left(y_{c}-y_{+-}\right)
\end{eqnarray*}
 After some algebra, the final transformed FP equation is
\begin{eqnarray*}
\frac{\partial Q}{\partial} & = & \chi\{\frac{\partial}{\partial z_{c+}}(n_{spin}+n_{c})+\frac{\partial}{\partial z_{c-}}(n_{spin}-n_{c})+\frac{\partial}{\partial y_{c+}}\left(n_{spin}+n_{c}\right)+\frac{\partial}{\partial y_{c-}}(n_{spin}-n_{c})\\
 &  & +\frac{\partial}{\partial z_{c+}}\frac{\partial}{\partial z_{c+}}-\frac{\partial}{\partial z_{c-}}\frac{\partial}{\partial z_{c-}}-\frac{\partial}{\partial y_{c+}}\frac{\partial}{\partial y_{c+}}+\frac{\partial}{\partial y_{c-}}\frac{\partial}{\partial y_{c-}}\}Q
\end{eqnarray*}
where $n_{spin}=n_{+}-n_{-}$ represents the spin of system $A$.
This equation has both diagonal positive and negative diffusion elements.
Here, we summarise that
\begin{eqnarray*}
z_{c+} & = & z_{c}+(z_{+}-z_{-})/2\\
z_{c-} & = & z_{c}-(z_{+}-z_{-})/2\\
y_{c+} & = & y_{c}+(y_{+}-y_{-})/2\\
y_{c-} & = & y_{c}-(y_{+}-y_{-})/2
\end{eqnarray*}
where $z_{c}$ and $y_{c}$ are variables pertaining to the meter
$C$, and $z_{\pm}$ and $y_{\pm}$ pertain to the system $A$. Importantly,
we note that
\[
n_{spin}=e^{z_{+}-y_{+}}-e^{z_{-}-y_{-}}
\]
so that in a \emph{full solution} of the FP equation, \emph{there
is a feedback from both the negative and positive diffusion terms
into the spin} $n_{spin}$.

However, the variables $n_{spin}$ and $n_{c}$ are, to a first approximation
in the limit of a macroscopic meter, constant. In this limit, assuming
constant $n_{spin}$ and $n_{c}$, we use the Theorem of the main
paper to write the exact stochastic equations as forward-backward
equations
\begin{eqnarray}
\frac{dz_{c+}}{dt} & = & -\chi\{n_{spin}+n_{c}\}+\eta_{z_{c+}}(t)\nonumber \\
{\color{black}{\color{red}{\normalcolor \frac{dz_{c-}}{dt_{-}}}}} & {\color{black}{\color{red}{\normalcolor =}}} & {\color{black}{\color{red}{\normalcolor -\chi\{n_{spin}-n_{c}\}+\eta_{z_{c-}}(t_{-})}}}\nonumber \\
{\normalcolor {\normalcolor {\normalcolor {\color{red}{\normalcolor \frac{dy_{c+}}{dt_{-}}}}}}} & {\normalcolor {\normalcolor {\color{red}{\normalcolor =}}}} & {\normalcolor {\color{red}{\normalcolor -\chi\{n_{spin}+n_{c}\}+\eta_{y_{c+}}(t_{-})}}}\nonumber \\
\frac{dy_{c-}}{dt} & = & -\chi\{n_{spin}-n_{c}\}+\eta_{y_{c-}}(t)\label{eq:solns-2}
\end{eqnarray}
Here, the first and fourth equations are solved in the standard way,
using an initial condition given by the variables defined by the Q
function at the initial time $t=0$. Here, nonzero correlations for
noise terms are $\left\langle \eta_{z_{c\pm}}\left(t\right)\eta_{z_{c\pm}}\left(t'\right)\right\rangle =2\chi\delta\left(t-t'\right)$,
$\left\langle \eta_{y_{c\pm}}\left(t\right)\eta_{y_{c\pm}}\left(t'\right)\right\rangle =2\chi\delta\left(t-t'\right)$.
The second and third equations arise from the terms with negative
diffusion and so need to be transformed into retrocausal equations.
Here, $t_{-}=-t$ denoted the backward time direction so that the
initial condition is determined by the future boundary condition i.e.
the Q function at the final time $T$. 

We have noted above that the diffusion terms are required for a consistent
quantum treatment of the meter. Looking at the FP equation, we can
say that $n_{spin}$ depends on $z_{+}-y_{+}$ (and $z_{-}-y_{-}$),
which are variables solved from $z_{c\pm}$ and $y_{c\pm}$, which
are a combination for the forward and backward propagation, involving
the noise terms. This gives a cyclic behaviour, where there is feedback,
and $n_{spin}$ depends on both future and past noise inputs.

In the approximation given by (\ref{eq:solns-2}), the noise terms
$\eta$ are included to solve for $z_{c\pm}$ and $y_{c\pm}$, while
keeping $n_{spin}$ constant. Here, the more complete equation for
$\dot{n}_{spin}$ is given by the above equation for $\dot{n}_{+}$
(and $\dot{n}_{-}$) but of type 
\begin{eqnarray*}
\dot{n}_{+} & = & \dot{\alpha_{+}}\alpha_{+}^{*}+\alpha_{+}\dot{\alpha_{+}^{*}}+noise=0+\frac{1}{|\gamma_{0}|^{2}}\widetilde{\eta}
\end{eqnarray*}
which includes noise terms $\widetilde{\eta}$. These give higher
order feedback into $n_{spin}$ (since $|\gamma_{0}|$ is large) which
has been ignored. We stress however that for a full treatment, $n_{spin}$
is not constant, and correction terms in the stochastic equations
(\ref{eq:solns-2}) may be necessary. 

\subsection{Solutions for the original coordinates}

To demonstrate the trajectories for the measurement, we focus on those
of meter mode, in particular the real part of $\alpha_{c}$ and of
the spin variable, $n_{spin}$, which is the constant of the motion
of the Hamiltonian interaction, identified with the spin that is being
measured. We demonstrate how the meter field $\alpha_{c}$ becomes
coupled to the spin-system, so that a final amplification of the meter
quadrature $X_{c}$ will give the spin value, as either ``up'' or
``down''.

Transforming back to the original coordinates, we see that for the
meter mode:
\begin{align*}
\alpha_{c} & =e^{\left(1+i\right)\left(z_{c}+iy_{c}\right)/2}=e^{(z_{c}-y_{c})/2}\{\cos\{(z_{c}+y_{c})/2\}+i\sin\{(z_{c}+y_{c})/2\}\}\\
n_{c} & =|\alpha_{c}|^{2}=e^{z_{c}-y_{c}}
\end{align*}
Here 
\begin{eqnarray*}
z_{c} & = & (z_{c+}+z_{c-})/2\\
y_{c} & = & (y_{c+}+y_{c-})/2
\end{eqnarray*}
which are solved from the above set (\ref{eq:solns-2}). The interesting
solution for the meter is the quadrature $X_{c}$ of the meter, since
this is the final measured amplitude that give the outcome for the
spin of $A$. The meter is initially decoupled from the spin system,
but at time $T$ there is a correlation so that positive $X_{c}$
values indicate spin ``up'', and negative $X$ values indicate spin
``down''. We identify the real part $\Re$ of variable $\alpha_{c}$
which is given by 
\[
\Re\alpha_{c}=e^{(z_{c}-y_{c})/2}\cos\{(z_{c}+y_{c})/2\}
\]
\emph{We see that the solution for the meter trajectories depends
on $z_{c}-y_{c}$ and $z_{c}+y_{c}$, which are combinations of both
backward and forward trajectories.} At any time $t$, the solution
for $\Re\alpha_{c}$ is given by adding or subtracting the solutions
for the forward and backward-solved variables.

The initial meter amplitudes are centred around $i\gamma_{0}$ the
amplitude of the original coherent state, the real part being zero.
For each simulation run, for the initial amplitude of a meter trajectory,
there is selected certain initial amplitudes $\alpha_{+}$ and $\alpha_{-}$
for the system fields, the meter and system being initially independent.
As $t$ increase toward the final time $T$, the correlation emerges:
For each run, the meter amplitude becomes real and diverges from zero
into one of two distinct macroscopic parts, either $\alpha_{c}\sim\gamma_{0}$
or $\alpha_{c}\sim-\gamma_{0}$. These represent the final spin outcomes
either ``up'' or ``down''. The values of $\alpha_{+}$ and $\alpha_{-}$
for the system conditioned on the final outcome $\alpha_{c}$ being
positive or negative form a set whose distribution is that of the
spin state ``up'' or ``down'' respectively. That this is so is
evident on examining the initial and final Q functions, given below.

\subsection{Q functions: initial and final states}

The boundary conditions for the forward-backward equations are determined
by the initial and final Q functions. These give the initial and final
probability distributions for the amplitudes, $\alpha_{c}$, $\alpha_{+}$
and $\alpha_{-}$, which are needed in the solving of the forward
and backward equations (\ref{eq:solns-2}).

The initial state of the spin system is the superposition of ``up''
and ``down'' eigenstates. 
\[
|\psi_{sup,s}\rangle=c_{1}|\uparrow\rangle_{a}+c_{2}|\downarrow\rangle_{a}=c_{1}|N\rangle_{a+}|0\rangle_{a-}+c_{2}|0\rangle_{a+}|N\rangle
\]
where we will take $N=1$. Including the independent meter in coherent
state $|\gamma\rangle$ ($\gamma=i\gamma_{0}$, where $\gamma_{0}$
is real), the Q function is defined as $Q(\alpha_{+},\alpha_{-},\alpha_{c})=\frac{|\langle\alpha_{+}|\langle\alpha_{-}|\langle\alpha_{c}|\psi\rangle|^{2}}{\pi^{3}}$
where $|\psi\rangle$ is the state of the system and meter. We find

\begin{eqnarray*}
Q_{initial} & = & \frac{1}{\pi^{3}}|\langle\alpha_{+}|\langle\alpha_{-}|\psi_{sup,s}\rangle\langle\alpha_{c}|\gamma\rangle|^{2}\\
 & = & \frac{e^{-|\alpha_{+}|^{2}-|\alpha_{-}|^{2}-|\alpha_{c}|^{2}-|\gamma|^{2}}}{N!\pi^{3}}|c_{1}\alpha_{+}^{*N}+c_{2}\alpha_{-}^{*N}|^{2}e^{\alpha_{c}^{*}\gamma+\alpha_{c}\gamma^{*}}
\end{eqnarray*}
This reduces to, taking $c_{1}$, $c_{2}$ real for convenience
\begin{eqnarray*}
Q_{initial} & = & \frac{1}{\pi^{3}}|\langle\alpha_{+}|\langle\alpha_{-}|\psi_{sup,s}\rangle\langle c|\gamma\rangle|^{2}\\
 & \rightarrow & \frac{e^{-(x_{A+}{}^{2}+p_{A+}^{2}+x_{A-}^{2}+p_{A-}^{2})-x_{C}^{2}-(p_{C}-|\gamma_{0}|)^{2}}}{\pi^{3}}\{|c_{1}|^{2}(x_{A+}^{2}+p_{A+}^{2})+|c_{2}|^{2}(x_{A-}^{2}+p_{A-}^{2})\\
 &  & +2c_{1}c_{2}(x_{A+}x_{A-}+p_{A+}p_{A-})\}
\end{eqnarray*}
where in the last line we have expressed the amplitudes in terms of
real and imaginary parts
\begin{eqnarray*}
\alpha_{+} & = & x_{A+}+ip_{A+}\\
\alpha_{-} & = & x_{A-}+ip_{A-}\\
\alpha_{c} & = & x_{c}+ip_{c}
\end{eqnarray*}

The final Q function for the entangled system-meter at time $T$ is
\begin{eqnarray*}
Q_{final} & = & \frac{1}{\pi^{3}}|\langle\alpha_{+}|\langle\alpha_{-}|\langle\alpha_{c}|\psi_{ent}\rangle|^{2}\\
 & = & \frac{e^{-|\alpha_{+}|^{2}-|\alpha_{-}|^{2}-|\alpha_{c}|^{2}-|\gamma_{0}|^{2}}}{N!\pi^{3}}|c_{1}e^{\alpha_{c}^{*}\gamma_{0}}\alpha_{+}^{N}+c_{2}e^{-\alpha_{c}^{*}\gamma_{0}}\alpha_{-}^{N}|^{2}
\end{eqnarray*}
which, on taking real and imaginary parts, reduces to (here $c_{1}=c_{2}=\frac{1}{\sqrt{2}}$)
\begin{eqnarray*}
Q_{final} & \rightarrow & \frac{e^{-(x_{A+}{}^{2}+p_{A+}^{2}+x_{A-}^{2}+p_{A-}^{2})-p_{C}^{2}}}{2\pi^{3}}\{e^{-(x_{c}-\gamma_{0})^{2}}(x_{A+}^{2}+p_{A+}^{2})+e^{-(x_{C}+\gamma_{0})^{2}}(x_{A-}^{2}+p_{A-}^{2})\\
 &  & +2\cos(p_{C}\gamma_{0})(x_{A-}x_{A+}+p_{A-}p_{A+})-2\sin(p_{C}\gamma_{0})(p_{A-}x_{A+}-x_{A-}p_{A+})\}
\end{eqnarray*}

\subsection{Final stage of measurement: ``collapse'' to the spin eigenstate}

The final stage of the spin measurement after the coupling of the
system to the meter is the measurement of $\hat{X}_{c}$, the quadrature
phase amplitude of the meter $C$ that is connected with the spin
value in the model $H_{I}$. 

The solutions for the outcome of $\hat{X}_{c}$ are given by the
amplitudes $x_{C}=\Re\alpha_{C}$ which are already amplified, and
can be amplified further by $H_{amp}$ of the main text. The equations
for $x_{C}$ decouple from those of $p_{C}$. Hence the initial conditions
for further amplification are determined by the marginal Q function
given by $Q_{final}$ integrated over\textbf{\emph{ $p_{C}$}}. This
gives
\begin{eqnarray*}
Q_{final} & = & \int dp_{C}Q_{final}\\
 & \rightarrow & \sqrt{\pi}\frac{e^{-(x_{A+}{}^{2}+p_{A+}^{2}+x_{A-}^{2}+p_{A-}^{2})}}{\pi^{3}}\{c_{1}^{2}e^{-(x_{C}-|\gamma_{0}|)^{2}}(x_{A+}^{2}+p_{A+}^{2})+c_{2}^{2}e^{-(x_{C}+|\gamma_{0}|)^{2}}(x_{A-}^{2}+p_{A-}^{2})\\
 &  & +2c_{1}c_{2}e^{-\gamma_{0}^{2}/4}(x_{A-}x_{A+}+p_{A-}p_{A+})\}
\end{eqnarray*}
For sufficiently large $\gamma_{0}$ (which is justified, the $\gamma_{0}$
being the initial \emph{macroscopic} amplitude of the meter), the
third term proportional to $e^{-\gamma_{0}^{2}}$ and which gives
interference will vanish. We see that in the simulation of $H_{I}$,
the final amplitudes for $x_{c}$ will be either $|\gamma_{0}|$ or
$-|\gamma_{0}|$. (There are two macroscopically distinct outcomes,
similar to the Figure 2a in the main text). The inferred state of
the system conditioned on the positive outcome $x_{C}=|\gamma_{0}|$
can be evaluated from the $Q_{final}$ and we see that the second
term that becomes proportional to $e^{-4|\gamma_{0}|^{2}}$will vanish,
but not the first term (for which the dependence on $\gamma_{0}$
vanishes). Hence, the inferred state associated with the positive
spin outcome $\gamma_{0}$ becomes precisely that of $|\uparrow\rangle$,
the spin eigenstate.

Similarly, the inferred state of the system conditioned on a negative
outcome $x_{C}=-|\gamma_{0}|$ for $x_{C}$ is $|\downarrow\rangle$
(the second term in brackets). This is regardless of the fact that
the Q functions of the $|\uparrow\rangle$ and $|\downarrow\rangle$
overlap. Thus, the meter quadrature $x_{c}$ is the important final
observable. The ``eigenvalue'' that gives the spin outcome of $+1$
(``up'') or$-1$ (``down'') is the value of the sign of the meter
amplitude $x_{C}=\Re\alpha_{c}$.

\subsection{Summary}

In summary, for each run of the simulation of $H_{I}$, the final
state of the meter system is such that the meter amplitude is real
and has diverged from zero into one of two distinct macroscopic parts,
either $\alpha_{c}\sim\gamma_{0}$ or $\alpha_{c}\sim-\gamma_{0}$.
These represent the final measured spin outcomes, either ``up''
or ``down''.

Following the procedure given in the main text and in the companion
paper and explained above, we can evaluate the inferred state for
the spin system, conditioned on a given outcome of the meter, either
$\alpha_{c}\sim\gamma_{0}$ or $\alpha_{c}\sim-\gamma_{0}$. \textbf{\emph{The
inferred state of the system associated with the positive spin outcome
$\gamma_{0}$ becomes precisely that of the spin eigenstate $|\uparrow\rangle$.}}
Similarly, the inferred state of the system conditioned on the negative
spin outcome $-\gamma_{0}$ is that of $|\downarrow\rangle$. The
values of $\alpha_{+}$ and $\alpha_{-}$ for the original spin system
conditioned on the final outcome $\alpha_{c}$ being positive or negative
form a set whose distribution is that of the spin state ``up'' or
``down'' respectively.

\section{Supplemental Materials: Violation of Bell inequalities using spin}

\subsection{Initial state}

The set-up is similar to that of Figure 4 of the main text, except
that there is an additional coupling to the meter at each site (Figure
\ref{fig:bell-2}). The common realisation of the Bell state is in
terms of four bosonic modes 
\[
|\psi_{bell}\rangle=\frac{1}{\sqrt{2}}(|1\rangle_{a+}|0\rangle_{a-}|1\rangle_{b+}|0\rangle_{b-}+|0\rangle_{a+}|1\rangle_{a-}|0\rangle_{b+}|1\rangle_{b-})
\]
Here, we use notation as in the previous section. Once the systems
$A$ and $B$ are prepared in the Bell state, the Bell experiment
involves unitary rotations on the each system $A$ and $B$. These
unitary rotations occur prior to the amplification part of the measurement,
which we will refer to as the pointer measurement. The unitary rotations
are associated with the choice of measurement setting $\theta$ or
$\phi$, and prepare the system in the appropriate basis. To study
this system, we evaluate the $Q$ function of the Bell state: 
\[
Q_{bell}(\vec{\alpha})=\frac{1}{2\pi^{4}}e^{-(|\alpha_{+}|^{2}+|\alpha_{-}|^{2}+|\beta_{+}|^{2}+|\beta_{-}|^{2})}|\alpha_{+}\beta_{+}+\alpha_{-}\beta_{-}|^{2}
\]
The Q function is a function of amplitudes $\vec{\alpha}=(\alpha_{+},\alpha_{-},\beta_{+},\beta_{-})$
each of which has a real and imaginary part.

\subsection{Measurement setting dynamics}

The polariser beam splitter determines the measurement setting: The
Hamiltonian is (for the setting at one site)
\[
H_{\theta}=\kappa(\hat{a}_{+}^{\dagger}\hat{a}_{-}+\hat{a}_{+}\hat{a}_{-}^{\dagger})
\]
The FP equation for the Q function is
\begin{eqnarray*}
\frac{dQ(\vec{\alpha})}{dt} & = & i\kappa(\frac{\partial}{\partial\alpha_{+}})\alpha_{-}Q+i\kappa(\frac{\partial}{\partial\alpha_{-}})\alpha_{+}Q-i\kappa(\frac{\partial}{\partial\alpha_{-}^{*}})\alpha_{+}^{*}Q-i\kappa(\frac{\partial}{\partial\alpha_{+}^{*}})\alpha_{-}^{*}Q
\end{eqnarray*}
which is deterministic. This gives a transformation to new variables
\begin{eqnarray*}
c_{+} & = & \alpha_{+}\cos\kappa t_{a}-i\alpha_{-}\sin\kappa t_{a}\\
c_{-} & = & \alpha_{-}\cos\kappa t_{a}-i\alpha_{+}\sin\kappa t_{a}
\end{eqnarray*}
where $t_{a}$ is the interaction time, in the Q function, equivalent
to changing to the measurement basis. We let $\theta=\kappa t_{a}$
for the interaction at $A$. The variables corresponds to the modes
$C_{\pm}$ outgoing from the beam splitter at $A$, for which the
transformation is
\begin{eqnarray*}
\hat{c}{}_{+} & = & \hat{a}_{+}\cos\kappa t_{a}-i\hat{a}_{-}\sin\kappa t_{a}\\
\hat{c}{}_{-} & = & \hat{a}_{-}\cos\kappa t_{a}-i\hat{a}_{+}\sin\kappa t_{a}
\end{eqnarray*}
where $\hat{c}_{\pm}$are the boson operators for modes $C_{\pm}$.

A similar independent interaction takes place at $B,$ denoted $H_{\phi}=\kappa(\hat{b}_{+}^{\dagger}\hat{b}_{-}+\hat{b}_{+}\hat{b}_{-}^{\dagger})$
where $\phi=\kappa t_{b}$. The transformed variables are
\begin{eqnarray*}
d_{+} & = & b_{+}\cos\kappa t_{b}-ib_{-}\sin\kappa t_{b}\\
d_{-} & = & b_{-}\cos\kappa t_{b}-ib_{+}\sin\kappa t_{b}
\end{eqnarray*}
and the operators $\hat{d}_{\pm}$ for outgoing modes $D_{\pm}$ at
$B$ are defined similarly. Hence, we define a rotated set of amplitudes
$\vec{\alpha}_{rot}=(c_{+},c_{-},d_{+},d_{-})$, and write the transformed
$Q$ function at the time $t_{1}$ after both rotations as $Q_{rot}(\vec{\alpha}_{rot})$.

\begin{figure}
\begin{centering}
\par\end{centering}
\begin{centering}
\includegraphics[width=0.85\columnwidth]{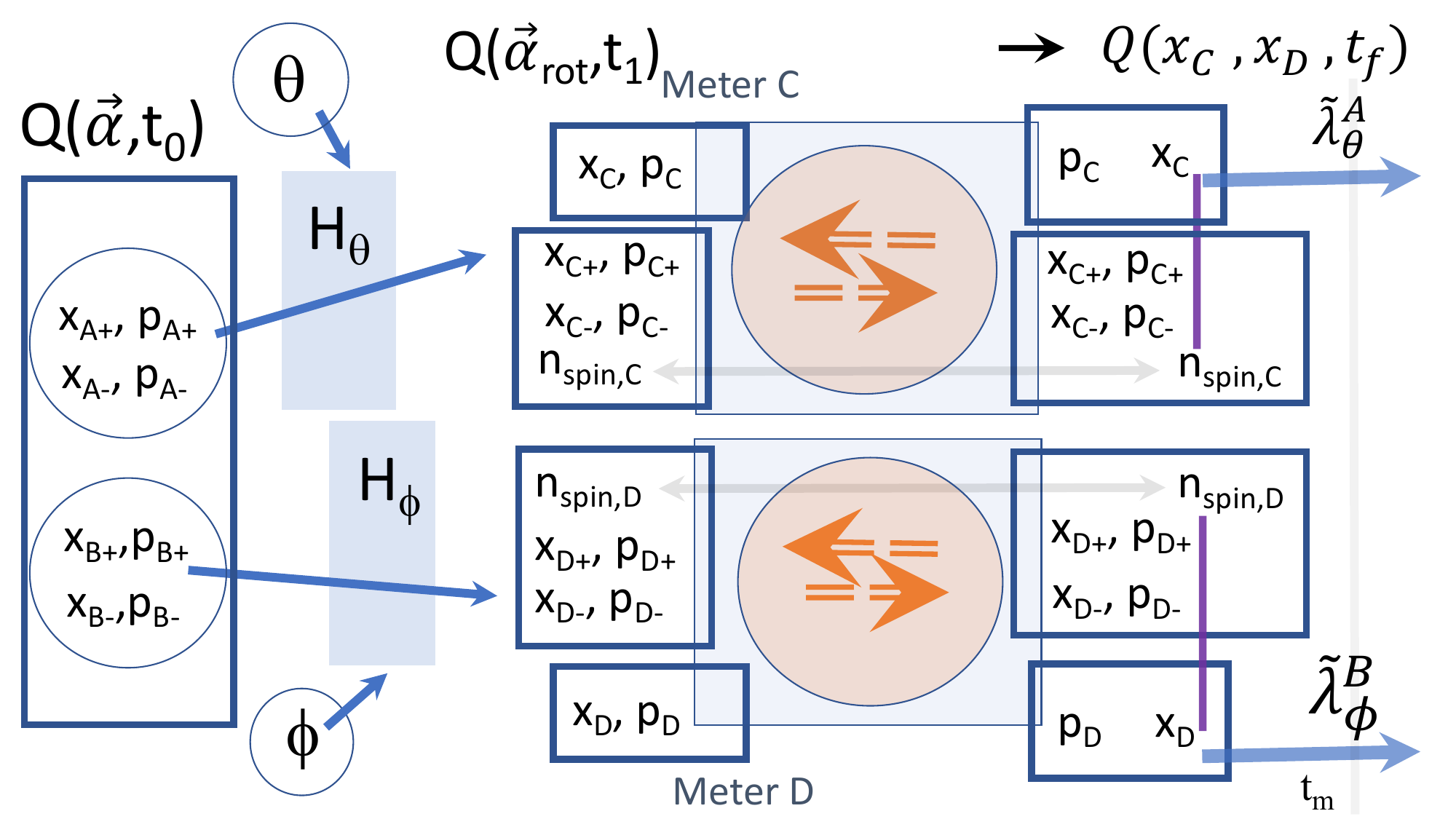}
\par\end{centering}
\caption{Schematic depiction of the simulation of the violation of Bell inequalities
based on spin measurements. The choice of measurement settings is
provided by interactions $H_{\theta}$ and $H_{\phi}$ at the separated
sites $A$ and $B$. These give local deterministic causal relations
between the amplitudes $x_{A\pm}$, $p_{A\pm}$ and $x_{C\pm}$ and
$p_{C\pm}$ (and between $x_{B\pm}$, $p_{B\pm}$ and $x_{D\pm}$
and $p_{D\pm}$). Each local system is then coupled to a meter, the
interaction being $H_{I}^{A}$ and $H_{I}^{B}$ as explained in the
previous section. The dynamics of the amplitudes is described by equations
given in the previous section, which involve forward-backward (and
potentially cyclic) causal behaviour. The system-meter couplings are
depicted by the orange circles in the blue boxes. The spin-variables
of each system (given by $n_{spin,C}$ and $n_{spin,D}$ which are
approximately constant) are coupled to the outgoing meter amplitudes
$x_{C}$ and $x_{D}$ respectively. The values of $n_{spin,C}$ and
$n_{spin,D}$ are not $\pm1$ however and do not correspond to the
``spin outcome''. The magnitudes of the meter amplitudes are large,
proportional to the initial amplitude $i\gamma_{0}$ of the meter.
The final ``spin outcomes'' correspond to the signs of the final
values of $x_{C}$ and $x_{D}$, which are either $\sim\gamma_{0}$
or $\sim-\gamma_{0}$. As the values are large (at the time $t_{m}$)
the ``element of reality'' spin variables $\widetilde{\lambda}_{\theta}^{A}$
and $\widetilde{\lambda}_{\phi}^{B}$ exist to predetermine the outcome
if there is a final detection at $t_{f}$. The trajectories conditioned
on the spin outcomes $\widetilde{\lambda}_{\theta}^{A}$ and $\widetilde{\lambda}_{\phi}^{B}$
being positive (for example) can be traced back to the time $t_{1}$,
and confirmed to correspond to the $Q$ function of the joint spin
eigenstate $|\uparrow\rangle_{A}|\uparrow\rangle_{B}$ in the $\theta$
and $\phi$ bases. Also, the trajectories conditioned on the spin
outcome $\widetilde{\lambda}_{\theta}^{A}$ can be traced back to
the original Bell state at time $t_{0}$, where the amplitudes for
$B$ for any future measurement $\sigma_{\phi}^{B}$ are fixed via
the original correlations (as in companion paper \citep{companion-paper}
and depicted by Figure 4b of the main text). This places a restriction
on the state for $B$, provided the settings at $A$ are fixed. \label{fig:bell-2}}
\end{figure}

\subsection{Spin measurements}

After the setting is fixed, the spins $\sigma_{\theta}^{A}$ and $\sigma_{\phi}^{B}$
defined at each site are measured. The measurement of the number difference
$n_{A}=\hat{c}_{+}^{\dagger}\hat{c}_{+}-\hat{c}_{-}^{\dagger}\hat{c}_{-}$
defines the spin outcome for $\hat{\sigma}_{\theta}^{A}$, and the
number difference $n_{B}=\hat{d}_{+}^{\dagger}\hat{d}_{+}-\hat{d}_{-}^{\dagger}\hat{d}_{-}$
defines the spin outcome for $\hat{\sigma}_{\phi}^{B}$.

The spins are measured using the local system-meter interactions $H_{I}^{A}$
and $H_{I}^{B}$, as defined above, at each site, where the modes
$A_{\pm}$ are now modes $C_{\pm}$, and the two modes at $B$ as
$D_{\pm}$. The two fields at each site are coupled to independent
meter systems, $C$ and $D$, the initial state of each meter being
a macroscopic coherent state $|i\gamma_{0}\rangle$. As explained
in Section 2, in the simulation the output amplitude $x_{\theta}^{C}$
of the meter $C$ at$A$ will be either at $\sim\gamma_{0}$ or $\sim-\gamma_{0}$,
which indicates the outcome of the spin measurement $\sigma_{\theta}$
as being $+1$ or $-1$ respectively. The output of the meter variable
$x_{\phi}^{D}$ at site $B$ will similarly reflect the outcome of
the spin $\sigma_{\phi}$ as being $+1$ or $-1$. The boundary condition
for the simulation is determined by the marginal $Q(x_{C},x_{D},t_{f})$
of the final Q function defined at time $t_{f}$ which includes the
meters (after the couplings via $H_{I}^{A}$ and $H_{I}^{B}$ which
are reversible). A depiction of the set-up is given in the Figure
\ref{fig:bell-2}. The set-up is similar to that depicted in Figure
4 of the main paper, with the amplitudes $x_{\theta}^{C}$ and $x_{\phi}^{D}$
giving the final outcome, except that the outcomes are binary.

The explanation of the Bell nonlocality is based on the retrocausal
feedback, as explained in the main text. Full details are given for
the CV case, in the companion paper \citep{companion-paper}.

\section{Supplemental Materials: Bell nonlocality}

\subsection{Breakdown of Bell's condition in the simulation model}

It is necessary to examine how the Bell nonlocality (defined as the
breakdown of Bell's local causality, or local hidden-variables, condition)
arises. The partial relaxation of EPR's local-realism with the variables
$\widetilde{\lambda}_{\theta}^{A}$, $\widetilde{\lambda}_{\phi}^{B}$
that determine a real property for the amplified system, is sufficient
to allow violations. The partial relaxation of the EPR's and Bell
premises is explained in the main text. Bell's local-hidden-variable
condition is 
\begin{equation}
P(++|\theta,\phi)=\int d\lambda\rho(\lambda)P_{A}(+|\lambda,\theta)P_{B}(+|\lambda,\phi)\label{eq:Bell-expression}
\end{equation}
where $P(++|\theta,\phi)$ is the probability of obtaining $+1$ at
both sites with settings $\theta$ and $\phi$; $\rho(\lambda)$ is
the distribution for hidden variables $\lambda$, and $P_{A}(+|\lambda,\theta)$
($P_{B}(+|\lambda,\phi)$) is the probability for $+1$ at $A$ ($B$),
given $\lambda$ and $\theta$ ($\phi)$. The condition refers to
measurements on a spin Bell state e.g.
\begin{equation}
|\psi_{bell}\rangle=\frac{1}{\sqrt{2}}\{|+\rangle_{A}|+\rangle_{B}+|-\rangle_{A}|-\rangle_{B}\}\label{eq:bell-state}
\end{equation}
where measurements of spins\textcolor{black}{{} $\hat{\sigma}_{\theta}^{A}$
and $\hat{\sigma}_{\phi}^{B}$ gives binary outcomes $\pm1$.} In
the case where quadrature phase amplitudes are measured, as referenced
in the main text, the Bell inequality is violated for certain states
where the $\pm1$ outcomes corresponding to the sign of $\hat{x}_{\theta}^{A}$
and $\hat{x}_{\phi}^{B}$.

The interactions $H_{\theta}^{A}$ and $H_{\phi}^{B}$ are local.
The question is to understand how the violation occurs if the system
is to be viewed as described by the hidden variables given by the
amplitudes of the Q function. This is explained in the companion paper
\citep{companion-paper}. The amplitudes $x_{\theta}^{A}$, $x_{\phi}^{B}$
of the simulation if directly measurable would lead to moments satisfying
Bell's condition, with $\rho(\lambda)\equiv Q(\bm{\lambda},t_{0})$.
However, parts of the amplitudes are ``hidden'', being associated
with noise inputs at the future boundary, and are not measured. The
Bell violations require changes of settings $\theta$ and $\phi$
at both sites, so that the Bell state is considered with respect to
rotations to two different bases, one for each system $A$ or $B$.
Over both unitary rotations, unobservable hidden terms {[}e.g. $\mathcal{I}$
in $Q_{bell}(\bm{\lambda},t_{0})${]} contribute to probabilities
for observable outcomes in the final $Q_{bell}(\bm{\lambda_{rot}},t_{1})$
{[}refer Figure 4 (top) of the main text{]}. These terms constitute
interference in the quantum state. Over the two changes of measurement
settings, the Bell condition breaks down:
\[
P(++|\theta,\phi)\neq\int d\lambda Q(\bm{\lambda},t_{0})P_{A}(+|\bm{\lambda},\theta)P_{B}(+|\bm{\lambda},\phi)
\]
 To give more detail, we summarise from the companion paper \citep{companion-paper}
and note that for a fixed choice of settings $\theta_{0}$ and $\phi_{0}$
at each site (at the time $t_{0}$) it is possible to define hidden
variables that predetermine the spin outcomes for the selected spin
measurements. We see this as follows: The system can be written in
the measurement basis as the superposition 
\begin{eqnarray}
|\psi_{bell}\rangle & = & c_{++}|+\rangle_{\theta_{0}}|+\rangle_{\phi_{0}}+c_{+-}|+\rangle_{\theta_{0}}|-\rangle_{\phi_{0}}+c_{-+}|+\rangle_{\theta_{0}}|-\rangle_{\phi_{0}}+c_{--}|-\rangle_{\theta_{0}}|-\rangle_{\phi_{0}}\nonumber \\
\label{eq:supms}
\end{eqnarray}
where the $c_{IJ}$ are probability amplitudes, and $IJ\in\{+,-\}$,
the states $|\pm\rangle_{\theta_{0}}$ being the Pauli spin eigenstates
of $\hat{\sigma}_{\theta_{0}}$ with eigenvalues $\pm1$ for the measurement
at $A$ (and $|\pm\rangle_{\phi_{0}}$ are defined similarly for the
spin measurement at $B$). The joint probabilities for the spin outcomes
are identical to those of the mixed state 
\begin{eqnarray*}
\rho_{mix} & = & \sum_{IJ}|c_{IJ}|^{2}\rho_{IJ}
\end{eqnarray*}
where $\rho_{IJ}=|I\rangle_{\theta_{0}}|J\rangle_{\phi_{0}}\langle J|_{\phi_{0}}\langle I|_{\theta_{0}}$,
for which the spin outcomes $I$ and $J$ occur with probability $|c_{IJ}|^{2}$
(refer Result below). Also, we see that for a given outcome, say $+1$
or $-1$, after amplification, it is possible to evaluate 
\[
P(x_{C+},x_{D+}|++)
\]
of $x_{C+}$ and $x_{D+}$ given the outcome $+1$ at both sites.
It is also possible to evaluate $P(++)$, the probability of outcome
$+1$ at both sites for the given $\theta_{0}$ and $\phi_{0}$, and
hence also the joint distribution $P(x_{C+},x_{D+},++)=P(x_{C+},x_{D+}|++)P(++)$.
Hence, we can evaluate 
\begin{eqnarray*}
P(+,+|x_{C+},x_{D+}) & = & P(x_{C+},x_{D+},++)P(x_{C+},x_{D+})\\
 & = & P(x_{C+},x_{D+}|++)P(++)P(x_{C+},x_{D+})
\end{eqnarray*}
 Next, we see (refer Result below) that if there is a change of setting
at \emph{one site only}, say from $\theta_{0}$ to $\theta$ at $A$,
then the probabilities for the spin outcomes are indistinguishable
from those of the mixed state obtained from a local transformation
on $\rho_{mix}$ (i.e. can be modelled by a local hidden variables
theory). However, if one considers the two Q functions, one for $|\psi_{bell}\rangle$
and one for $\rho_{mix}$, then there are however differences in the
``hidden'' terms $\mathcal{I}$ of the the Q functions. Allowing
for a further change of setting at $B$ from $\phi_{0}$ to $\phi$,
(i.e. with a change of both settings $\theta$ and $\phi$, after
time$t_{0}$), we see from the Result that the predictions for $|\psi_{bell}\rangle$
and $\rho_{mix}$ diverge. The contribution of the interference term
$\mathcal{I}$ to the probability for an observable event at $A$
after a change in setting from $\theta_{0}$ to $\theta$ is not necessarily
independent of the change of setting (from $\phi_{0}$ to $\phi$)
that may have occurred at $B$.

We explain the Bell nonlocality depicted in Figure 4 (top) of the
main text. The figure depicts a continuous variable (CV) Bell violation,
where the measurements are of the sign of the outcomes of quadrature
phase amplitude measurements. The system at the time $t_{1}$ has
been prepared with respect to the measurement basis, given by $\theta$
and $\phi$, and is described by the Q function $Q_{bell}(\bm{\lambda_{rot}},t_{1})$,
which depends on both settings. The Q function $Q_{bell}(\bm{\lambda_{rot}},t_{1})$
at time $t_{1}$ can be decomposed into sum of Gaussian distributions
with means given as the eigenvalues $x_{\theta,j}^{A}$ and $x_{\phi,k}^{B}$,
combined with interference terms, similar to that of eq (2) in the
main text. On amplification at each site, the means amplify to $Gx_{\theta,j}^{A}$
and $Gx_{\phi,j}^{B}$ to give detectable outcomes corresponding precisely
to the eigenvalues. The Q function written without the interference
terms is that of a mixed state $\rho_{mix,CV}$, for which the system
is probabilistically in a state with a definite outcome (that is one
of the eigenvalues). In the simulation, there is a forward deterministic
relation for the eigenvalue $x_{\theta,j}^{A}$ (the mean of the Gaussian
peaks) that leads to the variable $\widetilde{\lambda}_{\theta}^{A}$,
which corresponds to a \emph{band} of amplitudes, as described by
Figures 1 and 2 of the main text. This is depicted by the solid blue
line. The arrow hence also depicts that the outcomes \emph{given}
the fixed settings are as defined for those of the corresponding mixed
state $\rho_{mix,CV}$.

In summary, we have shown how a ``hidden variable'' model that has
causal properties different to Bell's model is able to display the
expected properties of satisfying Born's rule, while demonstrating
both EPR correlations and Bell violations, provided that a physical
model of the amplifying meter is included in the dynamics.

\subsection{Result: Nonlocality, settings, and the Q function}

\subsubsection*{No impact on $\widetilde{\lambda}_{x}$ at $A$ by a change of setting
at $B$}

We explain further the last statements in the main paper that justifies
the premise Condition (2). We follow the argument presented in the
companion paper and in Ref. \citep{weak}. Suppose the bipartite system
is prepared initially at time $t_{0}$ in, for example, the Bell state,
with $Q$ function $Q(\bm{\lambda},t_{0})$ as in Figure 4.

There is a device at each location that adjusts the measurement settings
$\theta$ and $\phi$, according to the interactions $H_{\theta}$
and $H_{\phi}$. We consider that the bipartite system comprised of
two systems, $A$ and $B$, is prepared at time $t=t_{0}$ for measurement
of $\hat{A}$ and $\hat{B}$ on the systems $A$ and $B$ respectively,
so that no adjustment of the measurement settings is necessary i.e.
we consider that the unitary operation associated with the choice
of measurement setting has occurred and the system is ready for coupling
to a meter/ and amplification. We can express the state in terms of
the prepared basis:
\begin{eqnarray}
|\psi_{ent}\rangle & = & \sum_{ij}c_{ij}|i\rangle_{A}|j\rangle_{B}\label{eq:bell-1}\\
 & \rightarrow & c_{11}|+\rangle|+\rangle+c_{12}|+\rangle|-\rangle+c_{21}|-\rangle|+\rangle+c_{22}|-\rangle|-\rangle\nonumber 
\end{eqnarray}
Here $|i\rangle$ and $|j\rangle$ are eigenstates of $\hat{A}$ and
$\hat{B}$, respectively, with eigenvalues $\lambda_{i}^{A}$ and
$\lambda_{j}^{B}$, and the $c_{ij}$ are probability amplitudes.
In the second line, we have expressed in terms of the spin eigenstates
relevant to the Bell state (\ref{eq:bell-state}). The Q function
$Q(\bm{\lambda},t_{0})$ is a sum of terms representing the Q function
of the eigenstates, plus interference terms $\mathcal{I}_{0}$.

We now change the measurement setting at site $B$ only. We hence
rotate the basis at $B$ by $\phi$, so that the system is prepared
for final coupling to a meter/ amplification to give the outcomes
for a \emph{new} measurement $\hat{B}_{\phi}$. Let the eigenstates
of $\hat{B}_{\phi}$ be $|k\rangle_{B}$ with eigenvalues $\lambda_{k}^{B}$.
Then we write $|j\rangle=\sum_{k}d_{jk}|k\rangle$. Thus, in the new
basis, state (\ref{eq:bell-1}) becomes
\begin{eqnarray*}
|\psi_{ent}\rangle_{\phi} & = & ==\sum_{ij}c_{ij}|i\rangle_{A}[\sum_{k}d_{jk}|k\rangle_{B}]=\sum_{i}\sum_{k}\sum_{j}c_{ij}d_{jk}|i\rangle_{A}|k\rangle_{B}=\sum_{i}\sum_{k}e_{ik}|i\rangle_{A}|k\rangle_{B}
\end{eqnarray*}
where $e_{ik}=\sum_{j}c_{ij}d_{jk}$. Looking at the example of the
Bell state, the Q function for this rotated state can be written as
a sum of the four new terms plus new interference terms $\mathcal{I}(\phi)$.
The joint probability for outcomes $i$ and $k$ is $|e_{ik}|^{2}=|\sum_{j}c_{ij}d_{jk}|^{2}=|c_{11}d_{11}+c_{12}d_{21}|^{2}+..$

We compare this result for $|\psi_{ent}\rangle$ with the system prepared
in a certain \emph{mixed} state $\rho_{mixA}$, in which $A$ can
be viewed as \emph{being in one} of the eigenstates $|i\rangle_{A}$
with a definite probability $|f_{i}|^{2}=\sum_{j}|c_{ij}|^{2}$. This
means that $A$ can be viewed as having a definite outcome $\widetilde{\lambda}^{A}$
for measurement $\hat{A}.$ We will show that after the change of
setting at $B$, there is no difference between the probabilities
for the outcomes of $\lambda_{i}$ and $\lambda_{k}$ of $|\psi_{ent}\rangle$
and $\rho_{mixA}$. The system is consistent with the interpretation
of having a definite value $\widetilde{\lambda}^{A}$ for the outcome
of $\hat{A}$ throughout an interaction $H_{\phi}$ giving a change
of setting at $B$.

To show this, we rewrite (\ref{eq:bell-1}) as 
\[
|\psi_{ent}\rangle=\sum_{i}f_{i}|i\rangle_{A}|\psi_{B}\rangle
\]
where $|\psi_{B}\rangle=\sum_{j}[c_{ij}/f_{i}]|j\rangle_{B}$. We
consider the mixed (non-entangled) state 
\[
\rho_{mixA}=\sum_{i}|f_{i}|^{2}|i\rangle|\psi_{B}\rangle\langle\psi_{B}|\langle i|
\]
The rotation $\phi$ at $B$ for the state $|\psi_{B}\rangle$ gives
\begin{eqnarray*}
|\psi_{B}\rangle_{\phi} & = & \sum_{j}\frac{c_{ij}}{f_{i}}\sum_{k}d_{jk}|k\rangle_{B}=\sum_{k}\sum_{j}\frac{c_{ij}}{f_{i}}d_{jk}|k\rangle_{B}\equiv\sum_{k}g_{k}|k\rangle_{B}
\end{eqnarray*}
Hence
\begin{eqnarray*}
\rho_{mixA,\phi} & = & \sum_{i}|f_{i}|^{2}\sum_{k,k'}g_{k}g_{k'}^{*}|i\rangle|k\rangle\langle i|\langle k'|
\end{eqnarray*}
For the Bell state example: 
\begin{eqnarray*}
\rho_{mixA,\phi} & \rightarrow & |c_{11}|^{2}|d_{11}|^{2}|+\rangle|+\rangle\langle+|\langle+|+|c_{11}|^{2}|d_{12}|^{2}|+\rangle|-\rangle\langle+|\langle-|\\
 &  & +|c_{22}|^{2}|d_{21}|^{2}|-\rangle|+\rangle\langle-|\langle+|+|c_{22}|^{2}|d_{22}|^{2}|-\rangle|-\rangle\langle-|\langle-|\\
 &  & +|c_{11}|^{2}\{d_{11}d_{12}^{*}|+\rangle|+\rangle\langle+|\langle-|+..
\end{eqnarray*}
The joint probability of outcomes $\lambda_{i}^{A}$ and $\lambda_{k}^{B}$
is 
\[
|f_{i}|^{2}|g_{k}|^{2}=|f_{i}|^{2}|\sum_{j}c_{ij}d_{jk}|^{2}/|f_{i}|^{2}=|\sum_{j}c_{ij}d_{jk}|^{2}
\]
in agreement with those of $|\psi_{ent}\rangle$.

What happens to the Q function, as the change of setting occurs at
$B$? The Q function for $\rho_{mix,\phi}$ includes interference
terms $\mathcal{I}_{mix}(\phi)$ (arising from off-diagonal elements
in the new basis) but the $|\psi_{ent}\rangle$ has extra interference
terms, since its density matrix has extra terms arising from the fact
that the system $A$ is in a superposition (i.e. originally entangled
with $B$). Hence, (even though the probabilities for (spin) outcomes
$\lambda_{i}^{A}$ and $\lambda_{k}^{B}$ for $|\psi_{ent}\rangle$
on rotation of basis at $B$ are the same as those of the system in
the mixed state $\rho_{mixA,\phi}$), there is a difference \emph{between
the interference terms }for the entangled state $|\psi_{ent}\rangle$
and the mixed state $\rho_{mixA,\phi}$.

\subsubsection*{Nonlocality becomes evident change of settings at $A$ and $B$}

This difference between the entangled state $|\psi_{ent}\rangle$
and $\rho_{mixA}$ (inherent in the interference terms which are ``hidden'')\emph{
becomes detectable on a} \emph{further change of setting} (i.e. rotation
of basis) at $A$. The $\rho_{mixA}$ is not entangled and does not
violate a Bell inequality. Consider a rotation $\theta$ at $A$ to
a new basis of eigenstates $|m\rangle$ so that the system is prepared
ready for a coupling to a meter/ amplification for measurement $\hat{A}_{\theta}$.
Taking the case of the Bell state e.g. $|\psi_{Bell}\rangle=\frac{1}{\sqrt{2}}(|+\rangle|+\rangle-|-\rangle|-\rangle)$,
the system after a single rotation at $B$ is of form
\begin{eqnarray*}
|\psi_{Bell}\rangle_{\phi} & \rightarrow & \frac{1}{\sqrt{2}}|+\rangle[\cos\frac{\phi}{2}|+\rangle+\sin\frac{\phi}{2}|-\rangle]-|-\rangle[\sin\frac{\phi}{2}|+\rangle-\cos\frac{\phi}{2}|-\rangle]
\end{eqnarray*}
where the probabilities for outcomes $++$, $+-$, $-+$, $--$ are
$\frac{1}{2}\cos^{2}\phi/2$, $\frac{1}{2}\sin^{2}\phi/2$, $\frac{1}{2}\sin^{2}\phi/2$
and $\frac{1}{2}\cos^{2}\phi/2$ respectively. The mixed state
\[
\rho_{mixA}=\frac{1}{2}\{|++\rangle\langle++|+|--\rangle\langle--|\}
\]
becomes on the first rotation 
\begin{eqnarray*}
\rho_{mixA,\phi} & = & \frac{1}{2}|+\rangle\langle+|\{\cos\frac{\phi}{2}|+\rangle+\sin\frac{\phi}{2}|-\rangle\}\{\cos\frac{\phi}{2}\langle+|+\sin\frac{\phi}{2}\langle-|\}\\
 &  & +\frac{1}{2}|-\rangle\langle-|\{\sin\frac{\phi}{2}|+\rangle-\cos\frac{\phi}{2}|-\rangle\}\{\sin\frac{\phi}{2}\langle+|-\cos\frac{\phi}{2}\langle-|\}
\end{eqnarray*}
which gives the same prediction for the probabilities as $|\psi_{bell}\rangle_{\phi}$.
However, after a rotation $\theta$, the Bell state is
\begin{eqnarray*}
|\psi_{bell}\rangle_{\phi,\theta} & = & \cos(\frac{\theta+\phi}{2})\{|+\rangle|+\rangle-|-\rangle|-\rangle\}+\sin(\frac{\theta+\phi}{2})\{|+\rangle|-\rangle+|-\rangle|+\rangle\}
\end{eqnarray*}
which gives probabilities $\cos^{2}[(\theta+\phi)/2]$ (and $\sin^{2}[(\theta+\phi)/2]$)
for outcomes $++$ and $--$ (and $+-$ and $-+$) respectively, which
are \emph{different} to those of the mixed state $\rho_{mixA}$. This
can be seen by evaluating $\rho_{mixA,\phi,\theta}$, defined by writing
$\rho_{mixA,\phi}$ in the new basis for $A$, but is evident by the
fact the the $|\psi_{bell}\rangle$ violates the Bell inequality,
whereas $\rho_{mixA}$ does not. The changes induced to the probabilities
after the second rotation are evident in the Q function $Q(\bm{\lambda},t_{1})$,
as the interference terms from the change of basis contribute to the
probability amplitudes for outcomes.

\end{widetext}


\begin{thebibliography}{99}
\bibitem{bell-1964}J. S. Bell, Physics \textbf{1}, 195 (1964).

\bibitem{bell-2004}J. S. Bell, Speakable and unspeakable in quantum
mechanics: Collected papers on quantum philosophy (Cambridge University
Press, 2004).

\bibitem{clauser-shimony-1978}J. F. Clauser and A. Shimony, Rep.
Prog. Phys. \textbf{41}, 1881 (1978). N. Brunner et al,  Rev. Mod.
Phys. \textbf{86}, 419 (2014).

\bibitem{eberhard-1989}P. H. Eberhard and R. Ross, Found. Phys. Lett.
\textbf{2}, 127 (1989).

\bibitem{pegg-1980}D. Pegg, Phys. Lett. A \textbf{78}, 233 (1980).
J. G. Cramer, Phys. Rev. D \textbf{22}, 362 (1980).

\bibitem{h-price-2008}H. Price, Studies in History and Philosophy
of Modern Physics \textbf{39}, 752 (2008). H. Price, Time's arrow
and Archimedes' point: new directions for the physics of time (Oxford
University Press, USA, 1996).

\bibitem{wharton-2020}K. B. Wharton and N. Argaman, Rev. Mod. Phys.
\textbf{92}, 021002 (2020). K. Wharton, Entropy \textbf{20}, 6 (2018).
N. Argaman, American Journal of Physics \textbf{78}, 1007 (2010).

\bibitem{donadi-hossenfelder}S. Donadi and S. Hossenfelder,  Phys.
Rev. A \textbf{106}, 022212 (2022).

\bibitem{araujo-2015}M. Araújo et al,  New Journ. Phys. \textbf{17},
102001 (2015).

\bibitem{barrett-2021}J. Barrett, R. Lorenz, and O. Oreshkov, Nature
communications \textbf{12}, 1 (2021). J.-M. A. Allen et al, Phys.
Rev. X \textbf{7}, 031021 (2017).

\bibitem{scully1982} R. Chaves, G. B. Lemos and J. Pienaar, Phys.
Rev. Lett. \textbf{120}, 190401 (2018).

\bibitem{time-order}G. Chiribella and Z. Lui, Communications Physics
\textbf{5}, 190 (2022).

\bibitem{giarmatzi-2019}C. Giarmatzi, in Rethinking Causality in
Quantum Mechanics (Springer, 2019) pp. 125--150.

\bibitem{costa-2016}F. Costa and S. Shrapnel, New J. Phys. \textbf{18},
063032 (2016). S. Shrapnel, The British Journal for the Philosophy
of Science \textbf{70}, 1 (2019).

\bibitem{drummond-2019}P. Drummond, Phys. Rev. Research \textbf{3},
013240 (2021).

\bibitem{hall-2020}M. Hall and C. Branciard, Phys. Rev. A \textbf{102},
052228 (2020).

\bibitem{hossenfelder-2020}S. Hossenfelder and T. Palmer, Front.
Phys. \textbf{8}, 139 (2020).

\bibitem{wood-2015}C. J. Wood and R. W. Spekkens, New J. Phys. \textbf{17},
033002 (2015). E. G. Cavalcanti, Phys. Rev. X \textbf{8}, 021018 (2018).
J. Pearl and E. G. Cavalcanti, Quantum \textbf{5}, 518 (2021).

\bibitem{finetuning}D. Almada, K. Ch'ng, S. Morrison and K.B. Wharton,
Int. J. Quant. Found. \textbf{2}, 1 (2016).

\bibitem{castagnoli-2021}G. Castagnoli, Phys. Rev. A \textbf{104},
032203 (2021).

\bibitem{vilasini-2021}V. Vilasini and R. Colbeck, Phys. Rev. A \textbf{106},
032204 (2022); Phys. Rev. Lett. \textbf{129}, 110401 (2022).

\bibitem{epr-1935}A. Einstein, B. Podolsky, and N. Rosen, Phys. Rev.
\textbf{47}, 777 (1935).

\bibitem{schrodinger-1935}E. Schrödinger, Naturwissenschaften \textbf{23},
\foreignlanguage{australian}{} 807-812, 823-828, 844-849 (1935).
P. Colciaghi et al, Phys. Rev. X\textbf{13}, 021031 (2023). 

\bibitem{leggett-1985}A. Leggett and A. Garg, Phys. Rev. Lett. \textbf{54},
857 (1985). \foreignlanguage{australian}{C. Emary, N. Lambert, and
F. Nori, Leggett-Garg inequalities, Rep. Prog. Phys \textbf{77},
016001 (2014).}

\bibitem{companion-paper}See companion paper arXiv 2205.06070 {[}quant-ph{]}
for details.

\bibitem{beables}\textcolor{black}{ J. S. Bell, ``Foundations of
Quantum Mechanics'', ed B d\textquoteright Espagnat (New York: Academic)
pp171-81 (1971).}

\bibitem{drummond-2020}P. D. Drummond and M. D. Reid, Phys. Rev.
Research\textbf{ 2}, 033266 (2020); Entropy \textbf{23}, 749 (2021).

\bibitem{simon-phi}S. Friederich, The British Journal for the Philosophy
of Science, 0, ja (2021), pp null. arXiv 2106.13502 (2021).

\bibitem{husimi-1940}K. Husimi, Proc. Phys. Math. Soc. Jpn. \textbf{22},
264 (1940).

\bibitem{Bohr} N. Bohr, Essays 1958-1962 on Atomic Phys. and Human
Knowledge vol. 3, (Ox Bow Press, Woodbridge, USA, 1987).

\bibitem{yuen-1976}H. P. Yuen, Phys. Rev. A \textbf{13}, 2226 (1976).

\bibitem{milburne-holmes-supQ}G. J. Milburn and C. A. Holmes, Phys.
Rev. Lett. \textbf{56}, 2237 (1986).

\bibitem{deutsch-1991}D. Deutsch, Phys. Rev. D \textbf{44}, 3197
(1991). M. Ringbauer, M.Broom, C. Myers, A. White and T. Ralph, Nature
Communications \textbf{5}, 1 (2014).

\bibitem{sm}Refer Supplemental Materials for details.

\bibitem{reid-1989}M. D. Reid, Phys. Rev. A \textbf{40}, 913 (1989).
M. D. Reid et al,  Rev. Mod. Phys. \textbf{81}, 1727 (2009).

\bibitem{gilchrist-prl-bell}U. Leonhardt and J. Vaccaro, Journ. Mod.
Opt.\textbf{ 42}, 939 (1995). A. Gilchrist, P. Deuar and M. D. Reid,
Phys. Rev. Lett.\textbf{ 80}, 3169 (1998); Phys. Rev. A \textbf{60},
4259 (1999).

\bibitem{weak}J. Fulton et al, Phys. Rev. A \textbf{110}, 022218
(2024);  J. Fulton et al, Entropy, \textbf{26}, 11 (2024); R. Rushin
Joseph et al, Phys. Rev. A\textbf{110}, 022219 (2024).\textcolor{black}{}
See also P. Grangier,  Entropy \textbf{23,} 1660 (2021)  O. J.
E. Maroney, Studies in History and Philosophy of Modern Physics,
\textbf{58}, 41 (2017).

\bibitem{macro-bell}M. Thenabadu and M. D. Reid,  Phys. Rev. A \textbf{105},
052207 (2022);  Phys. Rev. A \textbf{105}, 062209 (2022).

\bibitem{strict-tests}\textcolor{black}{M. Pusey, J. Barrett and
T. Rudolph, Nature. Phys. }\textbf{\textcolor{black}{8}}\textcolor{black}{,
475 (2012).} C. Brukner,  Entropy \textbf{20}, 350 (2018). D. Frauchiger
and R. Renner, Nat. Commun. \textbf{9}, 3711 (2018). K.W. Bong et
al,  Nature Phys. \textbf{16,} 1199 (2020). M. J. Hall,  Phys. Rev.
A \textbf{110}, 022209 (2024).
\end{thebibliography}
\end{document}